\documentclass[a4paper,fleqn,usenatbib]{mnras}

\usepackage{newtxtext,newtxmath}

\usepackage[T1]{fontenc}
\usepackage{ae,aecompl}

\usepackage{graphicx}	

\def\L{\mathcal{L}}

\title[Retrievals from high-resolution transit spectra]
{Relative abundance constraints from high-resolution optical transmission spectroscopy of WASP-121b, and a fast model-filtering technique for accelerating retrievals}

\author[N. P. Gibson et al.]{
Neale P. Gibson$^{1}$\thanks{E-mail: n.gibson@tcd.ie},
Stevanus K. Nugroho$^{2}$,
Joshua Lothringer$^{3}$,
Cathal Maguire$^{1}$,\newauthor
and David K. Sing$^{3,4}$
\smallskip
\\
$^{1}$School of Physics, Trinity College Dublin, University of Dublin, Dublin-2, Ireland\\
$^{2}$Astrobiology Center, NINS, 2-21-1 Osawa, Mitaka, Tokyo 181-8588, Japan\\
$^{3}$Department of Physics and Astronomy, Johns Hopkins University, Baltimore, MD, USA\\
$^{4}$Department of Earth and Planetary Sciences, Johns Hopkins University, Baltimore, MD, USA\\
}

\date{Accepted 2022 January 11. Received 2022 January 5; in original form 2021 October 19.}

\pubyear{2021}

\begin{document}
\label{firstpage}
\pagerange{\pageref{firstpage}--\pageref{lastpage}}
\maketitle

\begin{abstract}

High-resolution Doppler-resolved spectroscopy has presented new opportunities for studying the atmospheres of exoplanets. While the `classical' cross-correlation approach has proven to be efficient at finding atmospheric species, it is unable to perform direct atmospheric retrievals. Recent work has shown that retrievals are possible using a direct likelihood evaluation or likelihood `mappings'. The unique aspect of high-resolution retrievals is that the data-processing methods required to remove the stellar and telluric lines also distort the underlying exoplanet's signal and therefore the forward model must be pre-processed to match this filtering. This was the key remaining limitation in our previously published framework. This paper directly addresses this by introducing a simple and fast model-filtering technique that can replicate the processing performed by algorithms such as {\sc SysRem} and PCA. This enables retrievals to be performed without having to perform expensive injection and pre-processing steps for every model. We show that we can reliably constrain quantitative measures of the atmosphere from transmission spectra including the temperature-pressure profile, relative abundances, planetary velocities and rotational broadening parameters. Finally, we demonstrate our framework using UVES transmission spectroscopy of WASP-121b. We constrain the temperature-pressure profile and relative abundances of Fe, Cr, and V to be $\log_{10}(\chi_{\rm Fe}/\chi_{\rm Cr})$\,=\,1.66$\pm$0.28, $\log_{10}(\chi_{\rm Fe}/\chi_{\rm V})$\,=\,3.78$\pm$0.29 and $\log_{10}(\chi_{\rm Fe}/\chi_{\rm Mg})$\,=\,-1.26$\pm$0.60. The relative abundances are consistent with solar values, with the exception of Fe/Mg, where the large Mg abundance is probably explained by the escaping atmosphere of WASP-121b that is not accounted for in our atmospheric model.

\end{abstract}




\section{Introduction}

Transmission and emission spectroscopy are fundamental methods for the characterisation of transiting planets. These have traditionally been done using low-resolution, time-series observations, where the goal is to measure the fraction of starlight absorbed by the upper atmosphere (transmission), or alternatively obtain a direct spectrum of the planet (emission) after subtracting the stellar contribution. This remains extremely challenging, and relies on extremely stable time-series observations of transit, eclipses, and/or phase curves. Nonetheless, these methods recover a low-resolution spectrum of the planet, that can in turn be directly fitted with model atmospheres to obtain physical constraints of the atmospheric properties, e.g. the abundances or temperature-pressure profiles \citep[e.g.][]{2002ApJ...568..377C,2008MNRAS.385..109P,2016Natur.529...59S,2021MNRAS.503.4787W}. 

High-resolution Doppler-resolved spectroscopy has offered an alternative route to perform transmission or emission spectroscopy \citep[e.g.][]{2010Natur.465.1049S,2012Natur.486..502B,2012ApJ...753L..25R,2013MNRAS.436L..35B, 2017AJ....154..221N}. 
This approach uses the large Doppler-shift of the planet relative to the quasi-static stellar and telluric features to disentangle the planet's signal from its host star. Isolating the signal in velocity space does not rely on stable time-series observations, and therefore is more robust against systematic effects. This approach to detecting species in exoplanet atmospheres has flourished in recent years, with detections of multiple molecular, atomic and ionised species in a wide range of planetary atmospheres \citep[e.g.][]{2018Natur.560..453H,2020Natur.580..597E,2021AJ....162...73P, 2021ApJ...910L...9N,2021Natur.592..205G, 2021MNRAS.506.3853M}.

The difficulty with Doppler-resolved cross-correlation spectroscopy is that it does not offer an obvious route to extracting an exoplanet spectrum from the data, which can in turn be directly fitted with model atmospheres to perform retrievals, as in the case for low-resolution observations. Instead the standard method is to perform cross-correlation with a model template after pre-processing the time-series spectra to remove the stellar and telluric features, before integrating the signals over orbital phase. However, this approach does not allow for easy comparison of different model atmospheres. Recently, pioneering work by \citet{2019AJ....157..114B} demonstrated a way of performing full retrievals on high-resolution datasets by `mapping' the cross-correlation to a likelihood, showing that it is indeed possible to get quantitative constraints on atmospheric properties via high-resolution observations, and furthermore, that it is trivial (at least in principle) to couple information from low-resolution and high-resolution observations. More recently (and inspired by the work of \citealt{2019AJ....157..114B}), we introduced an alternative (but closely related) approach based on a direct evaluation of the likelihood \citep{2020MNRAS.493.2215G}. We note that there are many alternative methods to cross-correlation proposed to analyse and retrieve information from high-resolution exoplanet observations \citep[e.g.][]{2018A&A...619A...3P,2020AJ....159..192F,2019MNRAS.490.1991W,2022A&A...657A..23E}; however, this paper is focussed on direct likelihood evaluation which is (arguably) the simplest approach and easiest to interpret, as it is analogous to the approach used on low-resolution datasets.


Direct likelihood evaluation allows for forward models to be compared directly with the data, and therefore for full `retrievals' of atmospheric parameters. However, like all model-fitting, these methods will only work if the forward model is a good representation of the (noiseless) data. Furthermore, we cannot extract the exoplanet's spectrum to be separately fitted by a range of model atmospheres. Therefore, high-resolution models need to be fitted directly to the time-dependent, high-resolution spectra, {\it after} they have been pre-processed to remove the stellar and telluric signals. This means that we must pre-process the forward model in a similar way if we are to extract reliable information from high-resolution time-series.
In addition, we must also infer the noise properties of the data during the retrieval. In the case of low-resolution observations, this is taken into account 
during the fitting of light curves, and the noise properties of the exoplanet's spectrum are subsequently fixed. For high-resolution retrievals, we have the added complication of inferring the noise properties of the data {\it simultaneously} with the atmospheric model fitting. This makes the {\it statistical} framework more akin to fitting transit light curves, than for standard atmospheric retrievals.

This paper directly addresses the issue of model filtering in high-resolution retrievals. This issue has already been discussed at length in \citet{2019AJ....157..114B}. One proposed solution is to inject the forward model atmosphere into a representation of the telluric and stellar model, and re-process the forward model in the same way as the data. Indeed, this was demonstrated to work successfully both with airmass de-trending and also with PCA filtering \citep{2019AJ....157..114B,2021AJ....162...73P,2021Natur.598..580L}. However, this injection and re-filtering process is generally slow, and in many cases can become the bottleneck in each likelihood calculation. In addition, the framework we presented in \citet{2020MNRAS.493.2215G} used the {\sc SysRem} algorithm to filter the data, and it is not possible to perform {\sc SysRem} for each and every likelihood calculation. Instead, we filtered the 1D forward model atmosphere using a high-pass filter. While this provided a rough approximation to the removal of the planetary continuum (tested via injection/recovery tests), it is not accurate enough to perform truly robust retrievals from high-resolution datasets. 

Indeed, this was the major limitation of our retrieval framework, and this paper addresses this by introducing a general and fast filtering approach that can be applied where either PCA or SysRem has been used to pre-process the data. We validate its use by injection tests into real datasets and then demonstrate its use on UVES data of the ultra-hot Jupiter WASP-121b. WASP-121b has been the subject of multiple studies in recent years, finding a rich array of neutral and ionised metals as well as molecular features using both low- and high-resolution transit and emission spectroscopy \citep[e.g.][]{2016ApJ...822L...4E,
2017Natur.548...58E,
2019AJ....158...91S,
2019MNRAS.488.2222M,
2020A&A...635A.205B,
2020MNRAS.496.1638M,
2020MNRAS.493.2215G,
2020A&A...636A.117M,
2020A&A...641A.123H,
2020MNRAS.494..363C,
2021MNRAS.506.3853M,
2021MNRAS.503.4787W,
2021A&A...645A..24B,
2021ApJ...907L..47Y}, and therefore provides an excellent test case of our retrieval framework.

We first summarise the UVES data in Sect~\ref{sect:observations}, before summarising our retrieval framework in Sect~\ref{sect:analysis}, where we also introduce our forward model of transmission spectra and our filtering method. In Sects.~\ref{sect:injections} and \ref{sect:application} we then demonstrate our retrieval framework on injected datasets as well as on the WASP-121b UVES data. Finally, Sects.~\ref{sect:discussion} and \ref{sect:conclusion} present our results and conclusions.

\section{UVES observations of WASP-121b}
\label{sect:observations}

We use high-resolution transit observations of WASP-121b that have already been published in \citet{2020MNRAS.493.2215G}, \citet{2020A&A...636A.117M} and \citet{2021MNRAS.506.3853M}, and we only briefly recap them here. The observations were made using UVES \citep{2000SPIE.4008..534D} on the 8.2-m `Kueyen' telescope of the VLT (Unit Telescope 2). The dataset consists of a single transit, observed on the night of 2016 December 25 as part of program 098.C-0547 (PI: Gibson). We used an exposure time of 100 seconds (with a readout time of 24 seconds), taking 137 exposures over 4.7 hours, with approximately 83 during transit. Exposures 70--72 were clipped from the data analysis due to loss of guiding. The observations were made using dichroic \#2, resulting in observations being separated into red and blue `arms', where the \#2 and \#4 cross-dispersers were used, respectively. The blue arm used a `free template' with a central wavelength of 437\,nm, giving R$\sim$80,000 from $\approx$3,750 to 4,990\,\AA\ over 31 spectral orders. The red arm was observed using a standard setup with central wavelength of 680\,nm, giving wavelength coverage from roughly 4,940 to 8,700\,\AA\ over 54 spectral orders, with a resolution of R$\sim$110,000. However, not all of the red arm is useful, due to low S/N at the red end due to the throughput of the dichroic. Finally, the observations were performed using the image slicer (\#3) to maximise throughput, with a decker height of 10 arcseconds.

The data were analysed using a custom pipeline outlined in \citet{2019MNRAS.482..606G} and \citet{2020MNRAS.493.2215G}, which performed basic calibrations plus extraction of the spectral orders. We used simple aperture extraction, after using an optimal extraction algorithm to detect and replace outliers. The ESO/UVES pipeline was used to produce combined bias and flat-field frames, provide initial trace positions (central positions for the multiple slices), and perform the initial wavelength calibration. We do not perform background subtraction as the traces from the image slicer covers the length of the slit; however, this is later subtracted during the data pre-processing, and furthermore is negligible for our observations over most wavelength channels.

Due to small changes in the wavelength solution with time, we aligned the spectra by doppler-shifting (via linear interpolation) each spectrum to the wavelength grid of the first observation, after measuring the velocity shift using cross-correlation with a template spectrum. For the blue arm, we used a {\sc PHOENIX} stellar template that best matches WASP-121 \citep{2013A&A...553A...6H}. For the red arm, we used a telluric absorption spectrum generated using {\sc molecfit} \citep{2015A&A...576A..78K, 2015A&A...576A..77S}. This is because the stellar lines dominate the spectrum in the blue arm, and the telluric lines dominate the red arm, which therefore each provides more accurate wavelength corrections in the respective arms. Alignment with the stellar template automatically corrects for the barycentric velocity correction and systemic velocity of WASP-121. The red arm was additionally corrected for the barycentric velocity and systemic velocity after measurement of the wavelength alignment using the telluric template. Therefore all data are corrected to the rest frame of WASP-121.
Note that we ignore the impact of the the Rossiter McLaughlin and Doppler-shadow effects. This could have a small impact of efficiency of the pre-processing and systematics, but is negligible when compared to the planetary velocity as noted in previous studies \citep{2020MNRAS.493.2215G,2021MNRAS.506.3853M}\footnote{as both $v\sin i$ is small ($\approx$14\,km/s), and the orbit is close to polar ($\lambda$\,$\approx$\,260\,deg, see \citealt{2016MNRAS.458.4025D}), therefore {\sc SysRem} will efficiently remove any quasi-static features.}, so we expect little impact on our results.
We add that in the general case were these effects are significant, pre-processing the time-series spectra to account for them \citep[e.g.][]{2017A&A...603A..73Y,2020MNRAS.496..504N} is preferable within our likelihood framework, as dealing with the effects for every forward model would be computational expensive.
Similarly, we ignore the orbital velocity of the star, which is corrected for in the blue arm through the cross-correlation with stellar template, but not for the blue arm.

To optimise the information extracted from the transit, it is critical to have good initial estimates of the time- and wavelength-dependent noise. We followed \citet{2020MNRAS.493.2215G} and extracted estimates of the noise for each order by assuming that the noise is dominated by a Poisson term and a single background value, i.e. the standard deviation for each pixel takes the form $\sigma_i = \sqrt{aF_i + b}$, where $F_i$ is the measured flux for a given time and wavelength, and $a$ and $b$ are coefficients to be determined. We first obtain a representation of the noise in each order by subtracting a 5th-order Principal Component Analysis (PCA) model from the data to get a residual array $R_i$ (which approximately leaves a zero-mean array plus noise).
The coefficients $a$ and $b$ are found by fitting our noise model (i.e. $\sigma_i$ above) to each order, by optimising the likelihood (after removing constant terms):
\[
\ln\L(a,b) = -0.5\sum_i \left(\frac{R_i}{\sigma_i}\right)^2 - \sum_i\ln\sigma_i.
\]
At this point we also clean any excess outliers by replacement with the PCA model. Finally, we reconstructed the uncertainties array using the best-fit values for $a$ and $b$. As this array is in turn noisy, we then modelled it with another 5th-order PCA model, and this model becomes our final estimate of the noise\footnote{In principle we could use the same {\sc SysRem} model as is used for the final data analysis. However, this is not necessary as the goal here is only to get the approximate scaling of the time- and wavelength-dependent noise, and the absolute scaling is fitted directly in our retrievals. This is further discussed in Sect.~\ref{sect:injections}.}. This final step is required to remove any bias in the noise determination. This is due to a subtle effect caused by low values of $F_i$ always having lower uncertainties due to the assumption of Poisson noise, whereas in reality there is noise in the determination of the uncertainties. This can bias the fits when modelling residual datasets. See \citet{2020MNRAS.493.2215G} for further details, as well as a visualisation of the noise determination. 

\subsection{Pre-processing the spectra: removal of stellar and telluric features}
\label{sect:preprocessing}

As the filtering method is an important consideration in the retrieval framework, we briefly recap the methods used to pre-process and filter the data. We follow a common approach used in the literature for high-resolution cross-correlation \citep[e.g.][]{2010Natur.465.1049S,2012Natur.486..502B,2013MNRAS.436L..35B} in order to disentangle the continuously Doppler-shifting spectral lines of the exoplanet's atmosphere from the (effectively) static stellar and telluric lines.

First we apply a blaze correction following the procedure outlined in \citet{2019MNRAS.482..606G}. The procedure we apply does not remove the blaze function, but places every spectrum (for each order) on a `common' blaze. This accounts for time-dependent distortions of the instrument response, and improves the performance of the filtering methods. To achieve this, we divide each spectrum through by the median spectrum in each order, smooth each residual spectrum with a median and then Gaussian filter to get a blaze correction (relative to the median spectrum), and finally divide each of the original spectra through by its respective blaze correction.  Due to the instability of the blaze correction at the blue end of each order, we removed the first 600 pixels. We also clipped the last 60 pixels from each order. This left a total of 2,340 and 3,436 pixels in the blue and red arms, respectively. It is important for accurate retrievals that the blaze correction does not impact the underlying exoplanet signal, but only removes broad, smoothly varying changes in the blaze arising from systematic effects. For this reason, we used a much wider kernel width (501 pixels) and standard deviation (100 pixels) for the median and Gaussian filter than in our previous work. These values were determined from injection tests similar to those discussed in Sect.~\ref{sect:injections}, where we found the kernel width and standard deviation of 11 and 50 pixels, respectively, as used in \citet{2020MNRAS.493.2215G}, could significantly impact the retrievals by distorting the exoplanet signal\footnote{It is important to consider that seemingly `lightweight' corrections can have a large impact on the final retrievals, and it is important to test that the data processing steps do not impact the underlying exoplanet signal via injection tests.}.

The next step is to remove the stellar and telluric features from the data. To do this, similarly to \citet{2020MNRAS.493.2215G}, we use the {\sc SysRem} algorithm \citep{2005MNRAS.356.1466T}, which was first adapted for use on high-resolution datasets by \citet{2013MNRAS.436L..35B}. It can be seen as a generalisation of PCA that accounts for uncertainties in the data array, and has therefore become the method of choice for model-independent filtering of high-resolution data \citep[e.g.][]{2017AJ....153..138B,2017AJ....153..268E,2021ApJ...910L...9N}; however, more recent studies performing retrievals have reverted to basis-model fitting or PCA which can be applied more quickly within each likelihood calculation \citep[e.g.][]{2019AJ....157..114B,2021AJ....162...73P,2021Natur.598..580L}, whereas {\sc SysRem} is an iterative process and is much too slow. We later introduce in Sect.~\ref{sect:preprocessing_model} a general filtering method that can be applied either to PCA- or {\sc SysRem}-filtered data, where we also describe these algorithms in more detail; here, we briefly recap how we pre-process the data.

In \citet{2020MNRAS.493.2215G}, we constructed a model of each order by applying multiple passes of {\sc SysRem}, and then dividing through by the re-constructed model. This process preserves the relative line depths which is important for atmospheric retrievals. Here, we use a slightly different method. We first divide through by the median spectrum in each order, before applying {\sc SysRem} and subtracting the resultant model from the data. These methods are subtly different. While the first may more accurately preserve the line depths by dividing through by a more accurate systematics model, the latter method allows for a faster filtering process within each likelihood. Dividing through by the median spectra first is generally sufficient to properly normalise the data, providing {\sc SysRem} is not fitting for large scale variations in the stellar and telluric signals which could further modify relative line depths. Unless otherwise stated, we apply 15 passes of {\sc SysRem}. The uncertainties for each order determined earlier are divided through by the median spectrum to account for the pre-processing (subtraction of the {\sc SysRem} model does not modify the uncertainties).

\begin{figure}
\centering
\includegraphics[width=85mm]{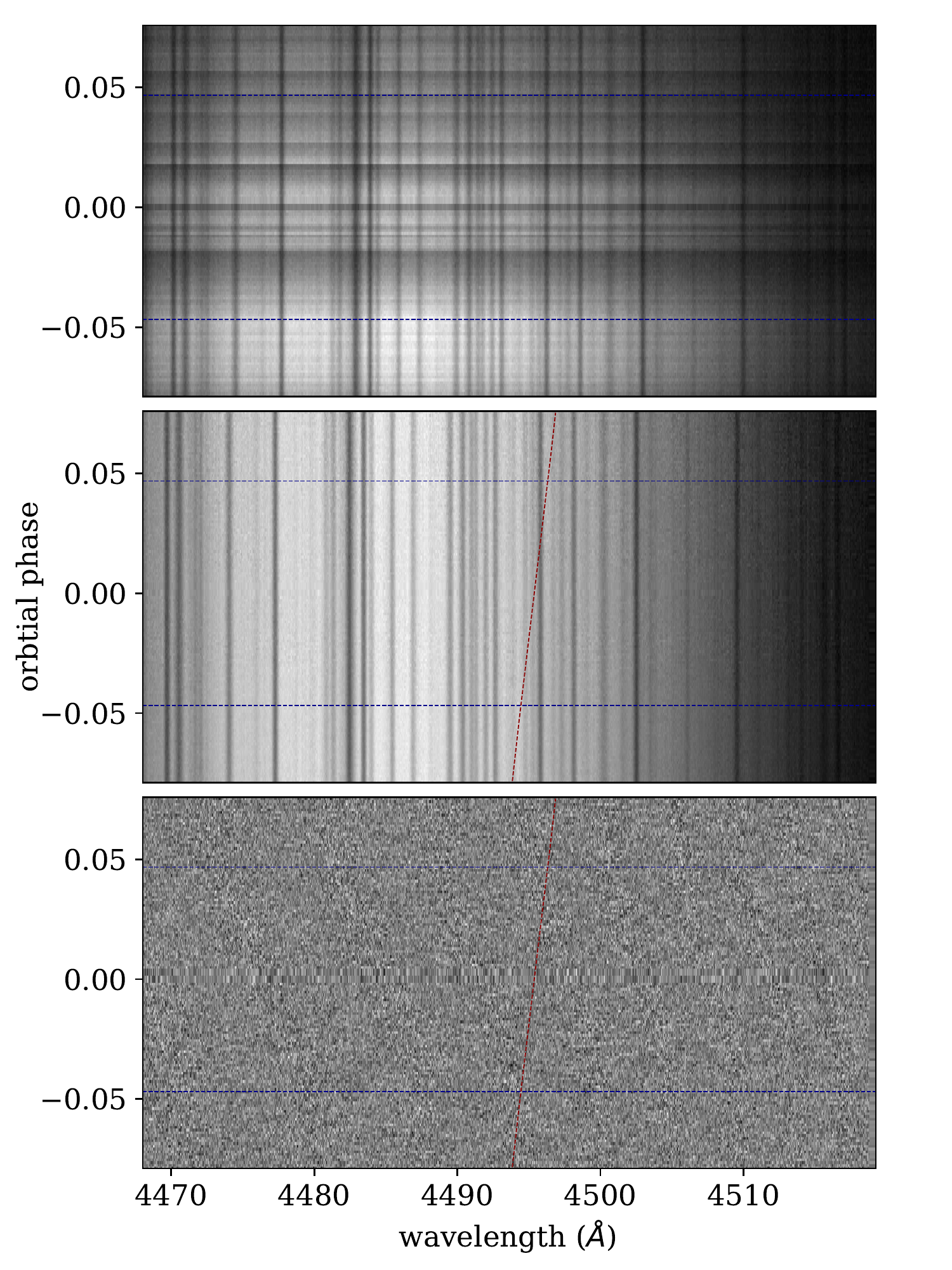}
\caption{Demonstration of the data pre-processing of a single order, including a raw order (top), after wavelength-alignment and blaze-correction (middle), and after division through by median spectrum and subtraction of the {\sc SysRem} model (bottom), here shown weighted by the uncertainties (standard deviations) to highlight the uniformity of the residual data. The horizontal dashed lines show the positions of transit ingress and egress, and the diagonal line shows the approximate velocity shift of WASP-121b.
}
\label{fig:processing}
\end{figure}

The standard pre-processing steps we apply are shown in Fig.~\ref{fig:processing} for a single order. As noted, the pre-processing will also modify the underlying exoplanet signal, and we must therefore account for this when fitting the data. Further discussion of PCA, {\sc SysRem}, and our filtering technique follows in Sect.~\ref{sect:preprocessing_model}.

%

\section{Likelihood Framework}
\label{sect:analysis}

We analyse our data using the likelihood approach outlined in \citet{2020MNRAS.493.2215G} in order to perform a full atmospheric retrieval on the data. Our method was inspired by \citet{2019AJ....157..114B}, and we also provided a more generalised from of their likelihood mapping that could account for time- and wavelength-dependent noise. We briefly discuss the similarities and differences in the derivations and results for various choices of likelihoods in Sects.~\ref{sect:likelihood_method} and ~\ref{sect:injections}.
In \citet{2020MNRAS.493.2215G}, we argued that out likelihood approach was `statistically rigorous'. However, the one remaining difficulty was how to account for the filtering of the exoplanet's signal with the application of {\sc SysRem} \citep[see Sect.~3.7 in][]{2020MNRAS.493.2215G}.

Our work used a simple filtering of the 1D forward model atmosphere (i.e. before Doppler shifting to account for the planet's motion); however, it was already clear from \citet{2019AJ....157..114B} that the filtering process needed to be applied to the 2D data arrays (time vs wavelength), i.e. the filtering is time-dependent.
More sophisticated methods based on PCA filtering have already been applied in the literature \citep{2021AJ....162...73P}. However, these methods generally involve injecting the Doppler-shifted atmospheric model into a noiseless approximation of the time vs wavelength array, and re-applying the filtering methods to approximate what is done to the underlying data. This is a slow process and a potential bottleneck in the retrieval process. In Sect.~\ref{sect:preprocessing_model} we outline how we sidestep this issue, and filter our models with a much faster approach that can also be applied to {\sc SysRem} filtering as well as PCA. However, we start by providing a brief recap of the cross-correlation and likelihood methods, and a description of our atmospheric model.

\subsection{Cross-correlation and the likelihood approach}
\label{sect:likelihood_method}

Both the cross-correlation and likelihood approach begin with a forward model of the planet's transmission spectrum at a given instant (our atmospheric model is discussed in detail in the following Section). In either case, the model is generated and will in turn be Doppler-shifted due to the planet's motion around its star given by:
\begin{equation}
\label{eq-velocities}
v_\mathrm{p} = K_\mathrm{p} \sin\,(2\pi\phi) + v_\mathrm{sys},
\end{equation}
where $K_\mathrm{p}$ is the velocity amplitude of the planet's orbit, generally assumed to be circular, $v_\mathrm{sys}$ is the systemic velocity, and $\phi$ is the orbital phase. In our case the data have already been shifted to the rest frame of the system, so the barycentric velocities can be safely ignored, and the systemic velocity becomes a measure of any additional redshift or blueshift relative to the system's rest frame.

In the `classical' cross-correlation approach, the model atmosphere typically consists of a single species plus continuum to create a fixed model template, which is then Doppler-shifted to a grid of systemic velocities ($v_\mathrm{sys}$). For each spectrum and value of $v_\mathrm{sys}$ the shifted model and data are multiplied and integrated over wavelength to produce a cross-correlation function (CCF). The data may first be weighted in various ways to account for the noise. In our case, we weight by the variance of the data, which is easily shown to be the optimal approach from the derivation of the likelihood below (we note however, that this assumes that we know the variance for each datapoint). The CCF, which is a function of $v_\mathrm{sys}$, can therefore be written as:
\begin{equation*}
\mathrm{CCF}\,(v_\mathrm{sys}) = \sum_{i}\frac{f_i m_i(v_\mathrm{sys})}{\sigma_i^2},
\end{equation*}
where $f_i$, $m_i$ and $\sigma_i$ are the (residual) data, model, and uncertainties for every time and wavelength, respectively, and the model is also dependent on $v_\mathrm{sys}$. The summation is generally first done over wavelength for each spectrum to produce a cross-correlation array of time (or phase) vs $v_\mathrm{sys}$ (for each spectral order).

The final step in the cross-correlation approach is to integrate the signal over time. As the planet's signal has a varying Doppler-shift with time, the cross-correlation functions are Doppler-shifted to the planet's rest frame before being integrated over time and order, after weighting on the transit function, to produce a final cross-correlation signal, again as a function of $v_\mathrm{sys}$. As we do not know the exact value for $K_\mathrm{p}$, we generally repeat this process over a grid of $K_\mathrm{p}$ values to produce a $K_\mathrm{p}$ vs $v_\mathrm{sys}$ `map'. This has the added advantage of visualising the impact of systematics across different values of $K_\mathrm{p}$.
Strong peaks in this map correspond to points in $K_\mathrm{p}$ and $v_\mathrm{sys}$ where the template model atmosphere can be found within the data. Where the peaks match the expected value for the planet's $K_\mathrm{p}$ and $v_\mathrm{sys}$, and there are no strong peaks elsewhere in the map (including negative `peaks'), we can be confident of detection of the atmospheric species contained within the template. In a nutshell, this is one of the major advantages of high-resolution Doppler-resolved spectroscopy over low-resolution methods; i.e. the fact that the signal can be separated in velocity space, where any systematic effects should be similar for different values of $K_\mathrm{p}$ and $v_\mathrm{sys}$, shows with high confidence that we are recovering a signal in the planet's atmosphere that correlates with the template model. There are a number of ways to determine the detection significance of the signal \citep[see e.g.][]{2010Natur.465.1049S,2013MNRAS.436L..35B}, but we do not go into detail here, other than to note that there is no statistically rigorous approach, and most methods compare the strength of the detected signal to the background in either the $K_\mathrm{p}$-$v_\mathrm{sys}$ map, or $\phi$-$v_\mathrm{sys}$ (vs order) arrays.

While the cross-correlation approach is extremely successful (and efficient) at finding atomic and molecular features in exoplanet atmospheres, it does not allow for direct comparisons between different model atmospheres, and therefore cannot be used for quantitative retrievals \citep[e.g.][]{2016ApJ...817..106B,2017ApJ...839L...2B}. A more statistically rigorous approach tries to compute the likelihood function rather than the CCF. See \citet{2019AJ....157..114B} for a detailed discussion, where the first rigorous framework for computing a likelihood `mapping' from the CCF for high-resolution datasets was introduced. Motivated by this work, \citet{2020MNRAS.493.2215G} provided an alternative approach using a full Gaussian likelihood function. The likelihood approach can either extend the classical cross-correlation approach to produce $K_\mathrm{p}$-$v_\mathrm{sys}$ maps of the likelihood rather than the CCF (which can then be directly compared for different models), or a single likelihood value can be computed for specific values for $K_\mathrm{p}$ and $v_\mathrm{sys}$ (by simply Doppler-shifting a forward model using these velocities), which can in turn be directly fed into optimisation functions or MCMC algorithms. The former approach can be seen as producing a grid of likelihood evaluations for a range of $K_\mathrm{p}$ and $v_\mathrm{sys}$, and remains extremely useful for visualising the signals and evaluating the impact of systematics.
The best approach depends on the number of free parameters in the forward model atmosphere, where grid searches become infeasible for large numbers of free parameters, therefore the latter is typically required for detailed atmospheric retrievals. Nonetheless, both approaches are mathematically equivalent.

To briefly recap the derivation from \citet{2020MNRAS.493.2215G}, we start with a full Gaussian (uncorrelated) likelihood:
\begin{equation*}
\L(\bmath\theta) = \prod_{i} \frac{1}{\sqrt{2\pi(\beta\sigma_i)^2}} \exp\left(-\frac{1}{2}\frac{(f_i-\alpha m_i)^2}{(\beta\sigma_i)^2} \right),
\end{equation*}
where $\alpha$ and $\beta$ are scale factors for the model and noise, respectively, and we have dropped the dependence of $m$ on $v_{\rm sys}$. $\bmath\theta$ represents a vector of model parameters, containing $v_{\rm sys}$, $K_{\rm p}$, $\alpha$, $\beta$, and any parameters of our forward model, $m$, and as before $i$ corresponds to each spectral order, wavelength, and time.

In practice we compute the log likelihood:
\begin{equation*}
\ln\L(\bmath\theta)  = -\frac{1}{2} \chi_\beta^2 -N\ln\beta-\sum_{i}\ln\sigma_i -\frac{N}{2}\ln2\pi,
\end{equation*}
where we define $\chi_\beta^{2}$ as:
\begin{equation*}
\chi^2 = \sum_i \frac{(f_i-\alpha m_i)^2}{(\beta\sigma_i)^2}.
\end{equation*}
Again, the summation over $i$ can be over wavelength, time, and/or order, depending on context. The final two terms in the log likelihood can be dropped as they are fixed, leaving:
\begin{equation}
\label{eq:full}
\ln\L (\bmath\theta)= -\frac{1}{2} \chi_\beta^2 -N\ln\beta.
\end{equation}
$\chi_\beta^{2}$ can be further expanded as:
\begin{equation}
\label{eq:chi2expansion}
\chi^2_\beta = \frac{1}{\beta^2}\left[\sum_i  \frac{f_i^2}{\sigma_i^2} + \alpha^2 \sum_i\frac{m_i^2}{\sigma_i^2} - 2\alpha \,{\rm CCF} \right].
\end{equation}
where the first summation term (within the brackets) is a constant and can be pre-computed, and the 2nd and 3rd summations are constant for a given set of model atmosphere parameters and velocities. Therefore the likelihood map may be calculated for a range of $\alpha$ and $\beta$ without having to recompute the CCF, which is a useful statistical trick\footnote{For example, we use the conditional distribution of $\alpha$ to compute the detection significances for each species which requires computation of only a single CCF, see Sect.~\ref{sect:application}.}. We use the combination of Eqs.~\ref{eq:full} and \ref{eq:chi2expansion} to compute both likelihood maps and single likelihood values, finding it to be the most computationally efficient method.

Following a similar `nulling' procedure to that described in \citet{2019AJ....157..114B} to remove the dependence of a single noise term $\sigma$, the dependence of $\beta$ can be removed from the likelihood, giving:
\begin{equation}
\label{eq:nulling}
\ln\L(\bmath\theta) = -\frac{N}{2}\ln \frac{\chi^{2}}{N},
\end{equation}
where $\chi^{2}$ no longer depends on $\beta$ and is defined as:
\begin{equation}
\chi^{2} = \sum_i \frac{(f_i-\alpha m_i)^2}{\sigma_i^2}.
\end{equation}
Note that this in effect optimises $\beta$ for a each given set of model parameters, so $\beta$ is no longer a property solely of the data, but also of the model.
This is a generalised version of the \citet{2019AJ....157..114B} likelihood mapping, that takes into account the time- and wavelength-dependent noise.
This becomes equivalent to the likelihood presented in \citet{2019AJ....157..114B} if we further drop the dependence on the uncertainties $\sigma_i$, and the scaling $\alpha$ (i.e. assuming that they both equal 1) to get:
\begin{equation}
\label{eq:nulling-noiseless}
\ln\L(\bmath\theta) = -\frac{N}{2}\ln \left[ \frac{1}{N}\left({\sum_i{f_i^2} + \sum_i{m_i^2} - 2 \sum_i{f_i m_i}} \right)\right].
\end{equation}
Dropping $\alpha$ in effect absorbs it into the forward model atmosphere, which depending on the forward model may result in an equivalent approach (e.g. a scale factor could be directly applied to the forward model $m$). 
Nonetheless, including it explicitly in the likelihood has some computational advantages as noted earlier. Indeed, some studies have already applied the \citet{2019AJ....157..114B} framework including $\alpha$ explicitly \citep{2021AJ....162...73P}.
However, the most significant difference between our approach and that of \citet{2019AJ....157..114B} is the addition of the time- and wavelength-dependence of the uncertainties.  \citet{2020MNRAS.493.2215G} report that the difference between Eqs.~\ref{eq:full} and \ref{eq:nulling} was found to be relatively insignificant, probably because the value of $\beta$ is well constrained using a large dataset. However, the difference between Eqs.~\ref{eq:full} or \ref{eq:nulling} vs \ref{eq:nulling-noiseless} is likely to be important for datasets with strong time- and wavelength-dependence of the noise.
Note that it is possible to partition the \citet{2019AJ....157..114B} log-likelihood into separate spectra, which in principle takes into account the time- (and order-) dependence of the noise \citep[e.g.][]{2021Natur.598..580L}, but it is difficult to account for the wavelength dependence in this way. It would similarly be possible to partition Eq.~\ref{eq:nulling} in this way, should the $\beta$ parameter be expected to change with time. 
Of course, it would also be possible to generalise Eq.~\ref{eq:full} to account for more complex noise parameters than a simple scaling.
We further explore this issue briefly in Sect.~\ref{sect:injections}. For full details of the derivations of these equations see \citet{2020MNRAS.493.2215G}.

Finally, we note that our UVES dataset for both blue and red arms contains 85 spectral orders, and 137 exposures, with 2,340 and 3,436 pixels in the blue and red orders, respectively (after clipping). Therefore we have more than 34 million data points to sum the likelihood over. Normally, computing a standard likelihood function is trivial; however, this huge number of data points highlights the need to be computationally efficient with all aspects of the calculation if we are to perform high-dimensionality retrievals using high-resolution data. This will become even more of an issue when we try and combine multiple datasets to explore time-dependence or use multiple instruments to explore broad spectral ranges. 

\subsection{Forward atmospheric model}
\label{sect:atmosphere}

The forward model atmosphere is of fundamental importance for the cross-correlation and likelihood methods, where we require an accurate template to match the underlying data and detect species, and/or make quantitative measurements. Within the likelihood framework, where we want to explore the posterior distribution using MCMC, tens or even hundreds of thousands of model templates need to be generated, making the atmospheric model a potential bottleneck in the calculations, particularly as they must be generated at high-resolution across a broad spectral range. Therefore it is critical that we can produce accurate forward models at speed. For this reason, in \citet{2020MNRAS.493.2215G} we used a simplified semi-analytic model, which assumed an isothermal atmosphere, and isobaric cross-sections (i.e. no pressure dependence). This was a modified version of the semi-analytic model of \citet{2017MNRAS.470.2972H}, using a different parameterisation and generalised to multiple atmospheric species \citep[also see][for an earlier approach using semi-analytic models]{2008A&A...481L..83L}.

In this work we use a more sophisticated model that takes into account the temperature-pressure profile and their impact on the absorption cross-sections, which is critical for recovering the temperature structure as well as reliable abundances. While there are now many open source codes to perform atmospheric forward modelling and retrievals \citep[e.g.][]{2019A&A...627A..67M,2017PASP..129d4402K,2013ApJ...779....3L, 2015ApJ...802..107W, 2021MNRAS.505.2675C}, we chose to develop a simplified transmission spectrum model that is designed to be as fast as possible, while being easily tuneable to the spectral ranges and resolution required by our data. Our model uses standard radiative transfer equations.

Our implementation, called `{\sc irradiator}', follows a similar approach to other transmission spectroscopy models. We first define a set of atmospheric layers covering a range of pressures (uniform in log space), and compute the $T$-$P$ profile using the parametric model from \citet{2010A&A...520A..27G} that has been adopted by many authors \citep[e.g.][]{
2013ApJ...779....3L,2019AJ....157..114B,2019A&A...627A..67M}. This requires four parameters; the irradiation temperature $T_\mathrm{irr}$, the mean infrared opacity $\kappa_{\rm IR}$, the ratio of visible-to-infrared opacity $\gamma$, and the internal temperature $T_\mathrm{int}$. Next, we compute the vertical extent (physical height) of the atmosphere assuming hydrostatic equilibrium, assuming a reference pressure, radius and gravity, and taking into account the varying temperature and gravity over the atmosphere. Rather than solve the equation directly in pressure, we solve this as a function of log-pressure. As the height varies more smoothly with the log of the pressure rather than with pressure, we find that this provides a more accurate numerical solution with fewer atmospheric layers.

With a temperature-pressure-altitude structure defined, we then compute the opacity of each layer in the atmosphere. Again, we follow a similar procedure to other transmission models, where we pre-define a set of absorption cross-sections from a pressure-temperature grid, and linearly interpolate each cross-section to the pressure and temperature of each atmospheric layer, weighting on the abundances of each species (with volume mixing ratios, VMRs, given by $\chi_{\rm species}$). In this work, we always assume a well-mixed atmosphere. We discuss the implications of this later. For each layer of the atmosphere, we then integrate through the grazing geometry of atmosphere, defining the sampling in each layer to coincide with the pressure-grid. Finally, we use the computed optical depth for each layer to convert to an effective radius. This is repeated for every point in the wavelength grid, which we typically compute with a resolution of R$\sim$200,000.

For the opacities, we focus on neutral atomic and singly ionised species which are known to dominate the atmosphere of WASP-121b at optical wavelengths \citep[e.g.][]{2020MNRAS.493.2215G,2020A&A...641A.123H,2021MNRAS.506.3853M}. In general, we use a range of opacity sources, including Kurucz atomic line lists \citet{2018ASPC..515...47K} and use the {\sc HELIOS-K} code to produce custom line lists \citep{2015ApJ...808..182G}. We also make use of the pre-computed opacity grids provided by petitRADTRANS\footnote{This is available at https://petitradtrans.readthedocs.io/en/latest/} \citep{2019A&A...627A..67M, 2020A&A...640A.131M} which allow us to test our implementation. For the remainder of this work, we use the pre-computed opacities provided by petitRADTRANS.

We use a simplified method to account for scattering. For Rayleigh scattering, we use the H$_2$ opacities from \citet{1962ApJ...136..690D}. We set the minimum allowed Rayleigh scattering to correspond to H$_2$ abundnace from Jupiter, and allow the effective H$_2$ volume mixing ratio ($\chi_{\rm ray}$) to increase freely to account for scattering from larger particles (where $\chi_{\rm ray}\approx0.855$ corresponds to pure H$_2$ scattering). Finally, we also include a parameterised cloud deck pressure ($P_\mathrm{cl}$), where we assume the atmosphere has infinite opacity below. To implement this in practice, we simply interpolate the cloud deck pressure to a cloud-deck altitude, then  truncate any values below the cloud deck radius after computing the radiative transfer.

To implement the radiative transfer, we write the equations as vector-matrix products were possible, in order to use the fastest possible implementations. We also store our cross-sections as single precision (32 bit) arrays rather than `standard' double precision (64 bit), as well as compute the full atmospheric model in single precision. The loss of precision is negligible, and far smaller than the loss of precision that occurs when interpolating cross-sections over a temperature and pressure grid (whether using single or double precision).
This immediately provides a factor of 2 improvement in terms of memory and even more in computational complexity, and in practice we get slightly better than a factor of 2 improvement in speed. We note that many open source codes compute the atmospheric model using double precision, and recommend the adoption of single precision to speed up calculations where rough approximations, e.g. interpolation over grids of cross sections, crude sampling of the T-P profile, already limit the precision of the model.

Our forward model atmosphere therefore has $N_{\rm species} + 6$ parameters, where $N_{\rm species}$ is the number of atmospheric species under consideration. The parameter vector is given by \{$T_\mathrm{irr}$, $\kappa_{\rm IR}$, $\gamma$, $T_\mathrm{int}$, $P_\mathrm{cl}$, $\chi_{\rm ray}$, $\chi_{\rm species}\times N_{\rm species}$\}.We used a reference pressure, radius and gravity of 0.01\,bar, 0.118\,$R_\star$ and 10.47\,m/s$^2$, respectively, to correspond to WASP-121b's atmosphere, a mean molecular weight of 2.33, and a stellar radius of $R_\star = 1.458 R_\odot$ \citep{2016MNRAS.458.4025D}. We do not adjust the mean molecular weight to correspond to abundances of the model. Constraining the chemistry using more sophisticated methods may allow better constraints on abundances, but will have limited impact on relative abundances. We briefly discuss this later. Furthermore, as we are allowing $\alpha$ to be a free parameter in the model fits, we are not especially sensitive to the values of the reference pressure.

\begin{figure*}
\centering
\includegraphics[width=150mm]{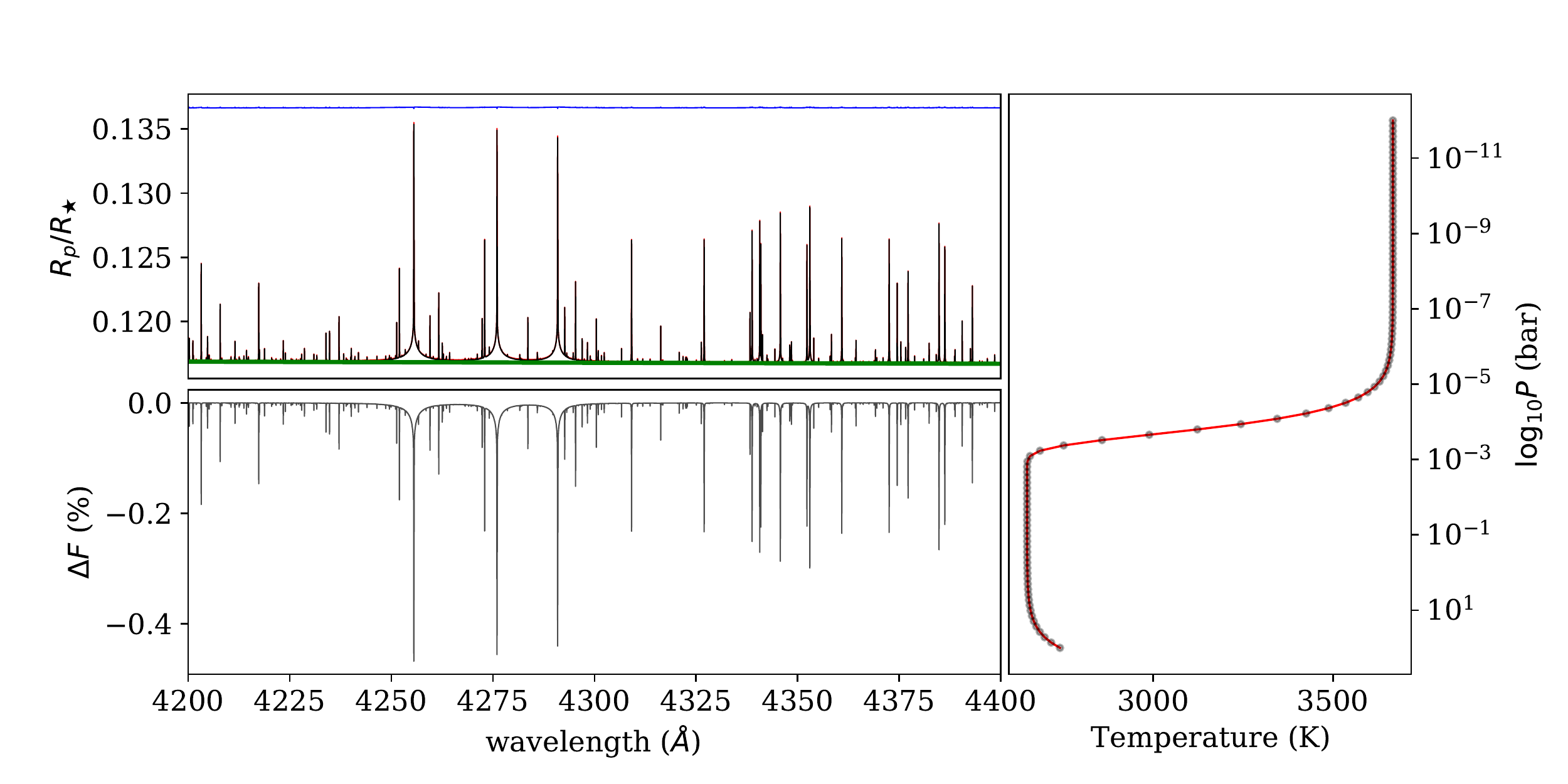}
\caption{Example of our forward model atmosphere, in units of planet-to-star radius ratio (upper-left) and relative flux (lower-left), and the corresponding temperature-pressure profile (right). In the upper left plot, the black line is our model, the green is the continuum, the red is the corresponding petitRADTRANS model (barely distinguishable), and the blue line is the difference between it and our model (with arbitrary offset). The points in the right plot correspond to the layers in the model atmosphere.}
\label{fig:forwardmodel1D}
\end{figure*}

Fig.~\ref{fig:forwardmodel1D} shows an example of our forward model atmosphere using atomic Fe and Cr with VMRs of 10$^{-8}$, H$_2$ Rayleigh scattering, and a cloud deck at 0.1\,bar. We used a model atmosphere with 100 layers, and an inverted T-P profile (also shown in the Figure).  We also computed a forward model with the same parameters using petitRADTRANS \citep{2019A&A...627A..67M, 2020A&A...640A.131M} in order to test our implementation, and the models are found to be consistent (see Fig.~\ref{fig:forwardmodel1D}). The very small differences are mainly due to the different method used to compute the altitude for the T-P grid from hydrostatic equilibrium (where as noted we computed it as a function of log pressure).

\subsection{Pre-processing the models to match the data}
\label{sect:preprocessing_model}

As discussed in Sect.~\ref{sect:preprocessing}, the data undergoes a series of pre-processing steps to remove the stellar and telluric features, as well as some artefacts from the instrument (e.g. blaze corrections, bad pixels). Critically, this will also alter the underlying exoplanet signal, by effectively filtering the exoplanet signal in a complicated way. The success of {\it any} data fitting or matching procedure relies on how well the forward model can represent the (noiseless) data. This is especially true of the likelihood method, where the purpose is to enable quantitative constraints on the exoplanet's atmospheric parameters.
For the detection of atomic and molecular species via cross-correlation, it is less critical if the forward model is an exact match. Generally speaking, as long as the lines are in the right place, and roughly the correct {\it relative} line strengths, it will still be efficient at finding a template match. Nonetheless, by improving the accuracy by which the forward model matches the (pre-processed) dataset, this should be able to boost the detection significances of atmospheric species, potentially important for marginal detections as well as to enable robust quantitative constraints on model atmospheric parameters.

As noted, this problem has already been discussed in the literature \citep[e.g.][]{2019AJ....157..114B}. In our previous work \citep[][]{2020MNRAS.493.2215G}, we used a simple high-pass filter on the (1D) forward model atmosphere to account for the removal of the continuum by {\sc SysRem}. However, it was clear that a filter applied to the 1D model could not accurately re-produce what {\sc SysRem} does to the underlying planetary signal, as the filtering will effectively change with time as the planet's signal is Doppler-shifted through the stellar and telluric spectrum. A more sophisticated approach is to apply the filtering to the Doppler-shifted datasets, which is the approach originally advocated by \citet{2019AJ....157..114B}, where a series of polynomial fitting is applied after injecting the model into a representative dataset. \citet{2021AJ....162...73P} adopt a similar technique, where the model is injected into a representative dataset and subsequently processed with PCA in order to replicate the filtering applied to the data. 

We suggest a new, faster filtering technique that can dramatically speed up each computation of the likelihood where the filtering technique is the bottleneck. This is true in our implementation, where the forward model, interpolation, and calculation of the likelihood (including our filtering) is orders of magnitude faster than {\sc SysRem} filtering, and significantly faster than PCA filtering. Generally, this will depend on the complexity of the forward model atmosphere.
Either way, this filtering enables the use of {\sc SysRem} rather than PCA, which is in principle a better way to filter the data as it naturally takes into account the noise in each datapoint. For a full description of {\sc SysRem}, see the original paper by \citet{2005MNRAS.356.1466T}, or for its application to high-resolution spectroscopy of exoplanets see \citet{2013MNRAS.436L..35B}. Here, we do not seek to give a full description of the algorithm, but just a conceptual overview that is necessary to outline our approach.

{\sc SysRem} is typically performed separately for each spectral order, and multiple `passes' of the algorithm are performed to remove common signals. 
For each pass of {\sc SysRem}, the time vs wavelength data array (which here we denote as $\mathbfss{A}$ for each spectral order) is decomposed into two column vectors, \mathbfit{u}  and \mathbfit{w}, where the model array for each pass ${p}$ is given by their outer product $\mathbfit{u}_{p}  \mathbfit{w}_{p}^{\rm T}$.
Here, \mathbfit{u} will have length equal to the number of spectra, and \mathbfit{w} will have length equal to the number of wavelength points in each order. The resulting model is then subtracted from the data to get the {\it processed} data after one pass, and the procedure is repeated on the processed data to get a new model array with a new set of \mathbfit{u} and \mathbfit{w}, which is again subtracted from the processed data. The final data array after $N$ passes of {\sc SysRem} then becomes:
\begin{equation*}
\mathbfss{A}_{N} = \mathbfss{A}_0 - \sum_{p=1}^N \mathbfit{u}_p  \mathbfit{w}_p^{\rm T} = \mathbfss{A}_0 - \mathbfss{U} \mathbfss{W}^{\rm T},
\end{equation*}
where
$\mathbfss{A}_0$ is the original array,
\mathbfss{U} and \mathbfss{W} are matrices containing all of the column vectors $\mathbfit{u}_p$ and $\mathbfit{w}_p$ (e.g. \mathbfss{U} is a K\,$\times$\,N matrix, where K is the number of spectra and N is the number of passes of {\sc SysRem}).

Each pass of the {\sc SysRem} algorithm requires an iterative procedure in order to determine the basis vectors. And while this is relatively efficient to run once on the whole dataset, it is very slow to run on the forward model for each and every likelihood evaluation. PCA can be considered as conceptually similar (where the noise is assumed constant for every datapoint), although it is much faster than {\sc SysRem} and is typically computed via singular value decomposition (although there may be faster algorithms to extract small numbers of basis vectors).

Our approach runs the filtering algorithm only once on the data, and uses the output matrix $\mathbfss{U}$ to perform filtering of the forward model within each likelihood calculation. The matrix multiplication $\mathbfss{U} \mathbfss{W}^{\rm T}$ can be considered as a simple linear basis model, where \mathbfss{U} contains the $N$ basis vectors, and \mathbfss{W} contains the corresponding weights. When we apply {\sc SysRem} or PCA to the data, we are fitting to {\it both} the stellar spectra and exoplanet spectra (as well as the tellurics plus systematics). It is therefore these basis vectors that are responsible for distorting the underlying exoplanet signal. While we are not able to decompose the \mathbfss{W} matrix into the weights that fit for the stellar and exoplanet signal, we can refit the basis weights for each of our forward models (this assumes the exoplanet signal can be decomposed linearly, which is a good assumption in this case where the exoplanet's signal is small). This process of fitting for the weights \mathbfss{W} using fixed \mathbfss{U} is faster than using either PCA or {\sc SysRem} to extract the new matrix decomposition for every likelihood evaluation. The intuitive reason is that these algorithms are trying to find the best fit \mathbfss{U} {\it and} \mathbfss{W} simultaneously; however, this extra computational complexity is not required as we have already extracted the correct basis models to filter the model.

Our approach is therefore to generate a forward model atmosphere, and doppler shift it to match the 3D data array (order vs time vs wavelength). For each order (time vs wavelength), we then fit the basis models \mathbfss{U} to the 2D model atmosphere for each order, hereafter \mathbfss{M} (note that each order has its own independent matrix of basis vectors \mathbfss{U}). We finally use this fitted array to perform a filtering over our forward models that closely corresponds to the distortions that {\sc SysRem} (or PCA) imposes on the underlying exoplanet signal.

In practice we use a simple linear least squares fitting to find the best-fit weights, e.g.:
\begin{equation*}
\bmath{w}^\prime = (\mathbfss{U}^{\rm T} \mathbfss{U})^{-1}\mathbfss{U}^{\rm T} \bmath{y}
\end{equation*}
in the case of a single column vector $\bmath{y}$, or:
\begin{equation*}
\mathbfss{W}^\prime  = (\mathbfss{U}^{\rm T} \mathbfss{U})^{-1}\mathbfss{U}^{\rm T} \mathbfss{Y}
\end{equation*}
for a 2D time vs wavelength array \mathbfss{Y}, i.e. this computes $N$ coefficients for each wavelength column in the array ($\mathbfss{W}^\prime$ is a L\,$\times$\,N matrix, where L is the number of wavelengths and N is the number of passes of {\sc SysRem}. Here, $\mathbfss{U}^\dagger = (\mathbfss{U}^{\rm T}\mathbfss{U})^{-1}\mathbfss{U}^{\rm T}$ is the Moore-Penrose inverse. To recover the best-fit model to the array \mathbfss{Y}, we simply have to multiply the derived weights by the basis vectors:
\begin{equation*}
\mathbfss{Y}^\prime  = \mathbfss{U} \mathbfss{U}^\dagger \mathbfss{Y}  .
\end{equation*}
Finally, to account for the data uncertainties in the fit (note that with {\sc SysRem}, these were already taken into account in the determination of \mathbfss{U}), we can perform a whitening transformation. To simplify this process and optimise the computation, we assume a single time-dependent array of uncertainties (for each order) by taking the mean of the uncertainties over wavelength $\hat\bsigma$. As the maximum likelihood estimator is not sensitive to the scaling of the uncertainties (and each wavelength channel is treated independently), we do not expect this to impact our results\footnote{While the uncertainties change with time and wavelength, this captures the major time-dependent trends for each wavelength channel.}.
Note that this is quite different to the complete likelihood calculation where we have emphasised that the time- and wavelength-dependence of the uncertainties are required to properly weight and combine the data. In addition, the correct scaling of the uncertainties (given by $\beta$) is critical to obtaining uncertainties in the retrieval.

Our final fits to the model in each order becomes:
\begin{equation}
\label{eq-filtering}
\mathbfss{M}^\prime  = \mathbfss{U}(\mathbf\Lambda\mathbfss{U})^\dagger (\mathbf\Lambda\mathbfss{M}),
\end{equation}
where $\mathbf\Lambda$ is a diagonal matrix of $1/\hat\bsigma$ terms. To account for the potential offset induced from dividing through by the median spectrum in each order (prior to application of {\sc SysRem}), we introduce a bias term to the basis vectors \mathbfss{U}, by simply introducing a new column filled with ones. Finally, in order to optimise the calculations, $\mathbfss{U}(\mathbf\Lambda\mathbfss{U})^\dagger\mathbf\Lambda$ can be precomputed as it only depends on the data and {\sc SysRem} basis vectors and is therefore fixed for every forward model.

The above procedure is used to perform a filtering of the forward model atmosphere within each likelihood computation. In practice we perform a number of steps to mimic the pre-processing described in Sect.~\ref{sect:preprocessing}.
We first generate the forward model atmosphere. Then, given values for $K_{\rm p}$ and $v_{\rm sys}$, we Doppler-shift the model as a function of time for each order (to match the 3D order vs time vs wavelength array), according to Eq.~\ref{eq-velocities}. We then weight each model according to the WASP-121b transit model, and add 1 to each spectrum (so the continuum is equal 1). In order to mimic the blaze correction, we then divide through by the median of each spectrum. Finally, we apply Eq.~\ref{eq-filtering} to the normalised data (independently for each order), and subtract the resulting model. The result is the 3D filtered version of the forward model that we can directly fit to the data.

\begin{figure*}
\centering
\includegraphics[width=180mm]{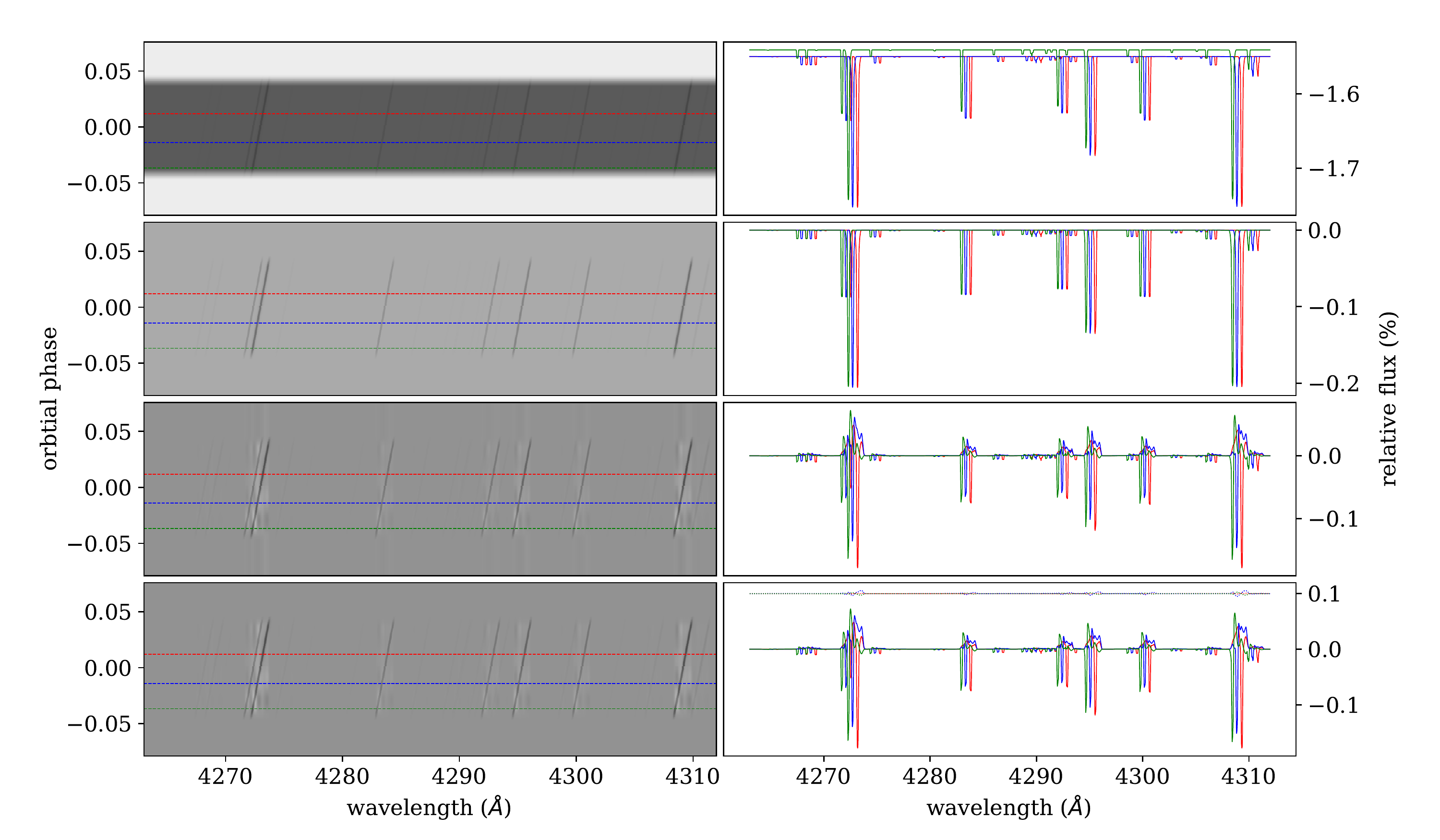}
\caption{Demonstration of the filtering technique applied to a single spectral order. Left: 2D time/phase vs wavelength array of the model for a single order. The top shows the Doppler-shifted forward model weighted on the transit. The panel below shows it after division through by the median of each spectrum (removing the transit shape). The third panel down shows the model after filtering, and is the final model fitted to the residual data set.
The bottom panel shows the same model injected into a noiseless representation of the stellar and telluric 2D array, after application of our standard pre-processing steps (including {\sc SysRem}), showing the extent of the distortion of the exoplanet's signal due to the pre-processing. The right panels show corresponding slices through the 2D arrays, at the locations indicated by the horizontal lines on the left. The similarity of the two bottom panels shows the effectiveness of our filtering method, where the dotted lines show the difference of the filtered models with the fully processed models (with arbitrary offset).}
\label{fig:model-processing}
\end{figure*}

Fig.~\ref{fig:model-processing} demonstrates these pre-processing steps on a single model atmosphere. We also compare this to a model injected into a noiseless representation of the stellar spectrum (see following Sect.), after pre-processing using our standard blaze correction plus {\sc SysRem}. The similarity of the resulting models shows that this method provides a satisfactory filtering of the model spectra, fully accounting for the time-dependence of the filtering. As noted, this process is also orders of magnitude faster than applying {\sc SysRem} directly for each model, and substantially faster than applying PCA, opening up the application of high-resolution retrievals to larger datasets and more complex atmospheric models.

\section{Application to Simulated and Real Data}

\subsection{Injections Tests}
\label{sect:injections}

In order to validate our filtering method (beyond a direct comparison of the forward models and the filtered, noiseless datasets), we run a series of injection tests and retrievals to ensure that we recover consistent model parameters. In order to keep the injection tests as realistic as possible, we inject our simulated signals directly into real time-series data, in this case the UVES WASP-121b transit data as described earlier. In order to speed up these injection tests, we use only 10 orders of the blue arm (orders 10--19 inclusive, roughly covering 4076--4476\,\AA). To balance the lower amount of data, we scale the amplitude of the injected models by 4--5 times the expected values. This is to provide a comparable precision to the real dataset (using all orders of the blue arm), but also tests the retrieval and filtering methods at larger amplitudes, where the filtering of the exoplanet atmosphere might be more extreme and more difficult to disentangle from the stellar filtering (as {\sc SysRem} will `target' the underlying exoplanet signal more if it is scaled up relative to the stellar signal).

Our process is similar to that used in \citet{2020MNRAS.493.2215G}, where we inject into the real data but using a negative velocity amplitude in order to ensure that the signal is well separated from the real exoplanet signal. This allows us to perform more realistic tests on noisy datasets. In order to prepare our simulated data, we first produce a 1D forward atmospheric model as described in Sect.~\ref{sect:atmosphere}. We compute the model with a constant resolution of R$\sim$200,000 (i.e. using a regular grid over log wavelength). In order to account for the fact that we do not know the exact broadening of the atmospheric features (e.g. due to rotation or winds), we then convolve the 1D model using a broadening kernel with width $W_{\rm conv}$. We experiment with both a Gaussian kernel (where $W_{\rm conv}$ is the standard deviation of the model sampling), and a uniform (`tophat') kernel (where $W_{\rm conv}$ is simply the kernel width). This allows for a constant velocity broadening of the model across the wavelength range (future work will consider more realistic broadening kernels for transmission spectra).
Strictly speaking, the resolution of the spectrograph is not constant, and changes both within and between spectral orders. To account for this, we could apply the broadening {\it after} the model is Doppler shifted to the wavelength grid of the data. However, this method is substantially slower than convolving the 1D model, and significantly slows down each likelihood calculation. The correct approach will also depend on whether the dominant broadening is astrophysical or instrumental. Another limitation is that if the 1D model is interpolated to coarser sampling, it is important that the line widths are broad in comparison to the data sampling, otherwise the use of interpolation (rather than binning) is invalid. Binning would be too slow to perform within each likelihood calculation, therefore broadening the 1D model prior to interpolation mitigates this potential problem.

After we have broadened our 1D model, we then interpolate it to the 3D model (order vs time vs wavelength) via linear interpolation (using Eq.~\ref{eq-velocities}) for a given set of $K_\mathrm{p}$ and $v_\mathrm{sys}$. We then weight the model according to the transit model of WASP-121b, and normalise so that the out-of-transit data equals 1, obtaining something similar to the upper-left panel of Fig.~\ref{fig:model-processing}. The final step is to inject this model into the data by multiplying into each order. We inject the model after the wavelength alignment, but before the blaze correction. We then perform the same pre-processing steps that are described in Sect.~\ref{sect:preprocessing} which includes the blaze correction and {\sc SysRem}. We use 5 passes of {\sc SysRem} for the injection tests, and store the basis vectors $\mathbfss{U}$. We then add a bias (constant) vector to $\mathbfss{U}$, and pre-compute $\mathbfss{U}(\mathbf\Lambda\mathbfss{U})^\dagger$, as well as the mean inverse variance as a function of time $\mathbf\Lambda$. This is done independently for each order. These fixed arrays are then ready for the filtering of the forward models.

We are now ready to perform our retrieval analysis on the injected data. To do this, we simply compute our forward model in the same way as the injected signal, taking into account the parameters of the atmospheric model \{$T_\mathrm{irr}$, $\kappa_{\rm ir}$, $\gamma$, $T_\mathrm{int}$, $P_\mathrm{cl}$, $\chi_{\rm ray}$, $\chi_{\rm species} \times N_{\rm species}$\}, the broadening kernel width \{$W_{\rm conv}$\}, the planet velocities \{$K_\mathrm{p}$, $v_\mathrm{sys}$\}, and the scale factors of the model and noise \{$\alpha$, $\beta$\}. In order to account for the scale factor $\alpha$ in the injection, we re-compute the atmospheric model with the scattering terms only, before subtracting from the full model and scaling by $\alpha$. We then add the continuum model back to the model before applying the interpolation and transit weighting. In the case of the retrieval, we can either subtract the continuum from each model, or leave it to be removed by the pre-processing\footnote{There could be a subtle difference due to this choice when normalising the models, where strictly speaking we should use the full model and normalise, but we find both methods work well in practice.}.

The next step for each forward model is to apply the model filtering as described in Sect.~\ref{sect:preprocessing_model}. We first divide each model spectrum (per order) by the median over wavelength. This imitates the effect of the blaze correction on each spectrum. We then fit each 2D array (time vs wavelength) using the basis models (i.e. using Eq.~\ref{eq-filtering}), subtracting this from the model fit. We are then left with a forward model for each order similar to that shown in the bottom panels of Fig.~\ref{fig:model-processing}. We finally compute the log posterior by computing the log prior then adding the log likelihood, i.e. the result of either Eqs.~\ref{eq:full}, \ref{eq:nulling} or \ref{eq:nulling-noiseless}. This is then fed into an MCMC algorithm to recover the posterior distributions of the model parameters. We use a Differential-Evolution Markov Chain \citep{2006S&C....16..239T,
2013PASP..125...83E}. In practice, we fit for the log of $\kappa_{\rm IR}$, $\gamma$, $P_\mathrm{cl}$, $\chi_{\rm ray}$, and $\chi_{\rm species}$, rather than fit for them directly. This does not impact the forward model, but does affect the (effective) priors placed on each of the model parameters.

\begin{figure*}
\centering
\includegraphics[width=150mm]{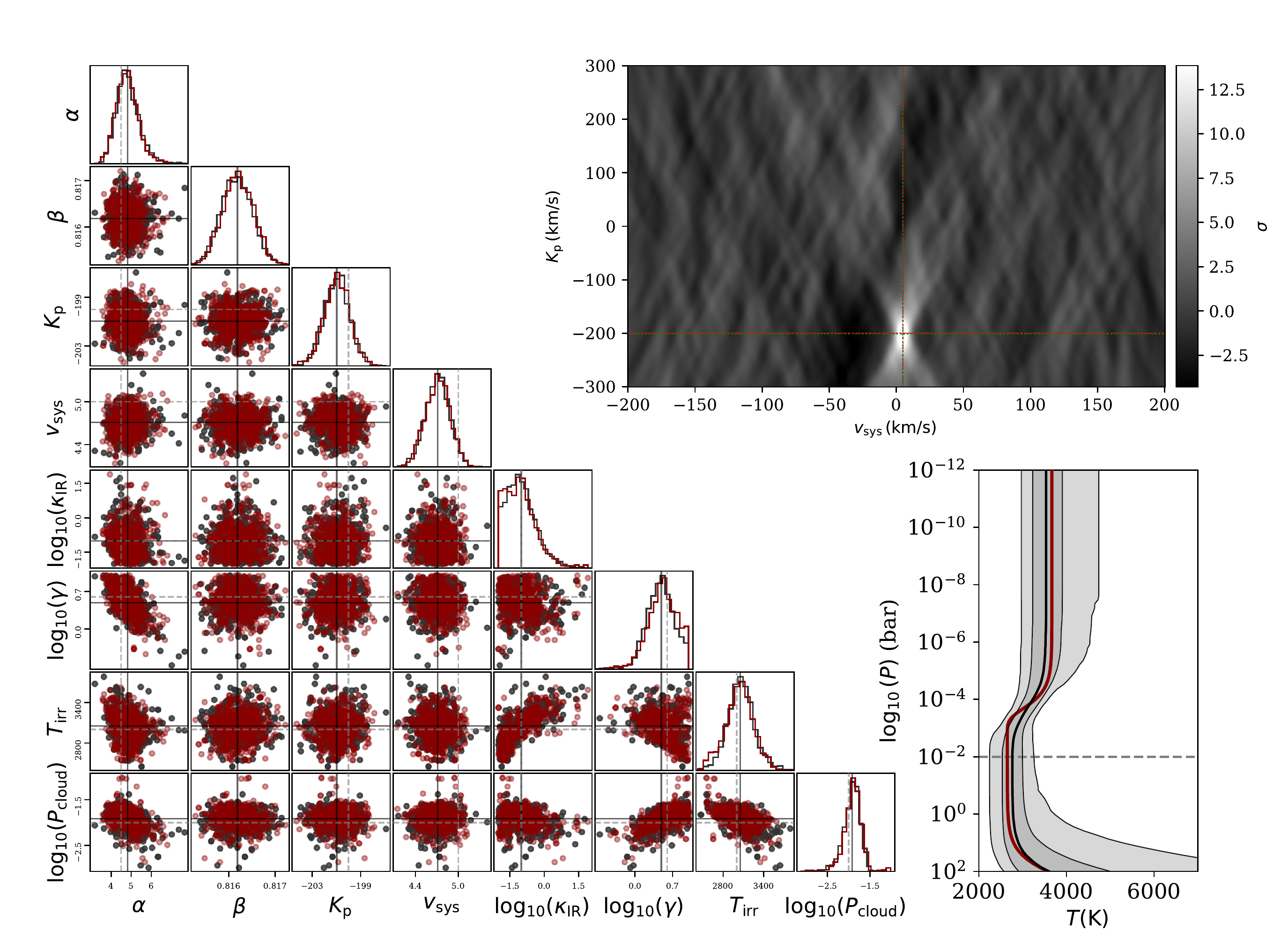}
\caption{Results of the first injection test. Left: 1D and 2D marginalised posterior distributions of the parameters from the MCMC fit. The different colours show the samples from two different sets of walkers. The injected and recovered values are shown by the dashed and solid lines, respectively. Upper right: Cross correlation `map' summed over orders. The dashed and dotted lines show the injected and recovered $K_{\rm p}$ and $v_\mathrm{sys}$, respectively. Bottom right: Temperature-pressure profile. The red line is the injected profile, and the grey shading marks the 1 and 2\,$\sigma$ recovered distribution, computed from 10,000 samples from the MCMC.}
\label{fig:injection1}
\end{figure*}

The parameters of our first injection test are shown in Tab.~\ref{tab:injectiontests}, where we inject a model atmosphere with only Fe plus scattering. For the first test, we fix the abundances, convolution width and Rayleigh scattering term, and also restrict the posterior using uniform priors (listed in the table). We run an MCMC chain with 128 walkers, with a burn-in length of 200 and chain length of 200, resulting in 51,200 samples of the posterior. We test for convergence using the Gelman \& Rubin statistic after splitting the chains into four separate groups, as well as checking the autocorrelation lengths.
For the first injection test, we use the full likelihood plus both versions of the `nulling' likelhoods\footnote{In the case of the \citet{2019AJ....157..114B} likelihood, we compute the log likelihood separately for each spectrum (and order), therefore accounting for the time- and order-dependence of the noise.}.
Fig.~\ref{fig:injection1} shows the results using the full likelihood (Eq.~\ref{eq:full}), including the 1D and 2D marginal distributions from the MCMC for each parameter, the temperature-pressure profile, and the $K_\mathrm{p}$-$v_\mathrm{sys}$ map of the likelihood (summed over all orders).
To visualise convergence, we divide the chains into groups of (independent) walkers, and overplot the 1D and 2D marginal distributions in different colours.
The temperature-pressure profile is computed for 10,000 random samples from the MCMC chains (using the values of $T_\mathrm{irr}$, $\kappa_{\rm IR}$, $\gamma$, and $T_\mathrm{int}$), and the median, and 1/2\,$\sigma$ ranges are computed from the resulting distribution.
The $K_\mathrm{p}$-$v_\mathrm{sys}$ map is computed directly from the CCF for a pre-defined grid of $v_\mathrm{sys}$ (1000 evenly-spaced samples from from -200 to 200\,km/s) and $K_\mathrm{p}$ (600 evenly-spaced samples from from -300 to 300\,km/s), using the filtered 3D model template for each order and time (using the best-fit $K_\mathrm{p}$ and $v_\mathrm{sys}$). The model is Doppler-shifted back to the stellar rest frame after performing the filtering before performing the cross-correlation as usual (but taking into account the time-dependence of the filtered model). The solid and dashed lines show the median recovered parameter and injected parameters, respectively, which are also presented in Tab.~\ref{tab:injectiontests}. Their consistency shows that the retrieved model parameters are robustly recovered using this framework, and that the filtering applied to each forward model fully accounts for the {\sc SysRem} filtering applied initially to the data. This is particularly evident in the value for the $\alpha$ parameter, where the data filtering is expected to significantly stretch or squash the spectral lines. The fact that we can confidently recover the model scaling shows that the filtering method can account for the distortions of the pre-processing on the real data.
The range of recovered temperature-pressure profiles is also shown (1 and 2\,$\sigma$), computed from the mean and standard deviation of 10,000 samples of the temperature-pressure parameters from the MCMC. The temperature and scattering properties are constrained by the relative line depths across the broad spectral range.

\begin{table}
\caption{Parameters for the two injection tests. $\mathcal{U}(a,b)$ represents a uniform (improper) prior, where the prior is zero if the parameter is outside the limits, and equal to one otherwise. See text for explanations.}
\label{tab:injectiontests}
\begin{tabular}{llll}
\hline
\noalign{\smallskip}
Parameter & Injected & Prior & Recovered\\ 
~[units] & value & ~ & value\\
\hline
\noalign{\smallskip}
\multicolumn{4}{l}{\it Injection Test 1:} \\\noalign{\smallskip}
~$\alpha$             & $4.5$ & $\mathcal{U}(0.1,8)$ & $4.79^{+0.57}_{-0.51}$  \\\noalign{\smallskip}
~$\beta$              & $1$ & $\mathcal{U}(0.1,2)$ & $0.8162\pm{0.0003}$ \\\noalign{\smallskip}
~$K_{\rm p}$ [km/s]          & $-200$ & $\mathcal{U}(-220,-180)$ & $-200.9\pm{1.1}$  \\\noalign{\smallskip}
~$v_{\rm sys}$ [km/s] & $5$ & $\mathcal{U}(-20,20)$ & $4.72^{+0.16}_{-0.17}$  \\\noalign{\smallskip}
~$W_{\rm conv}$      & $5$ & - & fixed \\\noalign{\smallskip}
~$\log_{10}(\kappa_{\rm IR})$ [m$^2$/kg] & $-1$ & $\mathcal{U}(-2,2)$ & $-1.08^{+0.70}_{-0.60}$  \\\noalign{\smallskip}
~$\log_{10}(\gamma)$  & $0.6021$ & $\mathcal{U}(-2,1)$ & $0.50\pm{0.27}$ \\\noalign{\smallskip}
~$T_{\rm irr}$ [K]        & $3000$ & $\mathcal{U}(2000,4500)$ & $3050^{+190}_{-200}$ \\\noalign{\smallskip}
~$T_{\rm int}$ [K]       & $200$ & - & fixed \\\noalign{\smallskip}
~$\log_{10}(P_{\rm cl})$ [bar] & $-2$ & $\mathcal{U}(-4,1)$ & $-1.90^{+0.16}_{-0.21}$ \\\noalign{\smallskip}
~$\log_{10}(\chi_{\rm ray})$ & $-0.068$ & - & fixed \\\noalign{\smallskip}
~$\log_{10}(\chi_{\rm Fe})$ & $-6$ & - & fixed \\\noalign{\smallskip}
\hline
\multicolumn{4}{l}{\it Injection Test 2:} \\\noalign{\smallskip}
~$\alpha$                       & $    3.0$ & $\mathcal{U}(0.1,8)$ & $3.20^{+0.41}_{-0.44}$ \\\noalign{\smallskip}
~$\beta$                        & $    1$ & $\mathcal{U}(0.1,2)$ & $0.8174\pm0.0003$ \\\noalign{\smallskip}
~$K_{\rm p}$ [km/s]                    & $ -200$ & $\mathcal{U}(-220,-180)$ & $     -200.3^{+    1.9}_{- 2.0}$ \\\noalign{\smallskip}
~$v_{\rm sys}$ [km/s]                  & $    5$ & $\mathcal{U}(-20,20)$ & $        4.48^{+    0.30}_{- 0.29}$ \\\noalign{\smallskip}
~$W_{\rm conv}$                 & $    2.5$ & $\mathcal{U}(1,50)$ & $      2.30\pm0.21$ \\\noalign{\smallskip}
~$\log_{10}(\kappa_{\rm IR}) $ [m$^2$/kg]   & $   -2.40$ & $\mathcal{U}(-4,0)$ & $     -2.63^{+    1.18}_{-0.97}$ \\\noalign{\smallskip}
~$\log_{10}(\gamma)$            & $    0.40$ & $\mathcal{U}(-2,1)$ & $          0.41^{+    0.53}_{-    0.45}$ \\\noalign{\smallskip}
~$T_{\rm irr}$ [K]                 & $ 3000$ & $\mathcal{U}(2000,4500)$ & $    3160^{+  450}_{-630}$ \\\noalign{\smallskip}
~$T_{\rm int}$  [K]               & $  200$ &       -              & fixed \\\noalign{\smallskip}
~$\log_{10}(P_{\rm cloud})$ [bar]    & $   -2$ & $\mathcal{U}(-4,1)$ & $       -3.34^{+0.7}_{- 0.46}$ \\\noalign{\smallskip}
~$\log_{10}(\chi_{\rm ray})$        & $   -0.068$ & - & fixed \\\noalign{\smallskip}
~$\log_{10}(\chi_{\rm Fe})$         & $   -6$ & $\mathcal{U}(-10,4)$ & $-5.41^{+0.60}_{-0.72}$ \\\noalign{\smallskip}
~$\log_{10}(\chi_{\rm Cr})$         & $   -7$ & $\mathcal{U}(-10,4)$ & $-6.71^{+0.58}_{-0.75}$\\\noalign{\smallskip}
~$\log_{10}(\chi_{\rm Mg})$         & $   -6$ & $\mathcal{U}(-10,4)$ & $-6.00^{+1.02}_{-1.65}$ \\\noalign{\smallskip}
\hline
\noalign{\smallskip}
\end{tabular}
\end{table}

\begin{figure*}
\centering
\includegraphics[width=175mm]{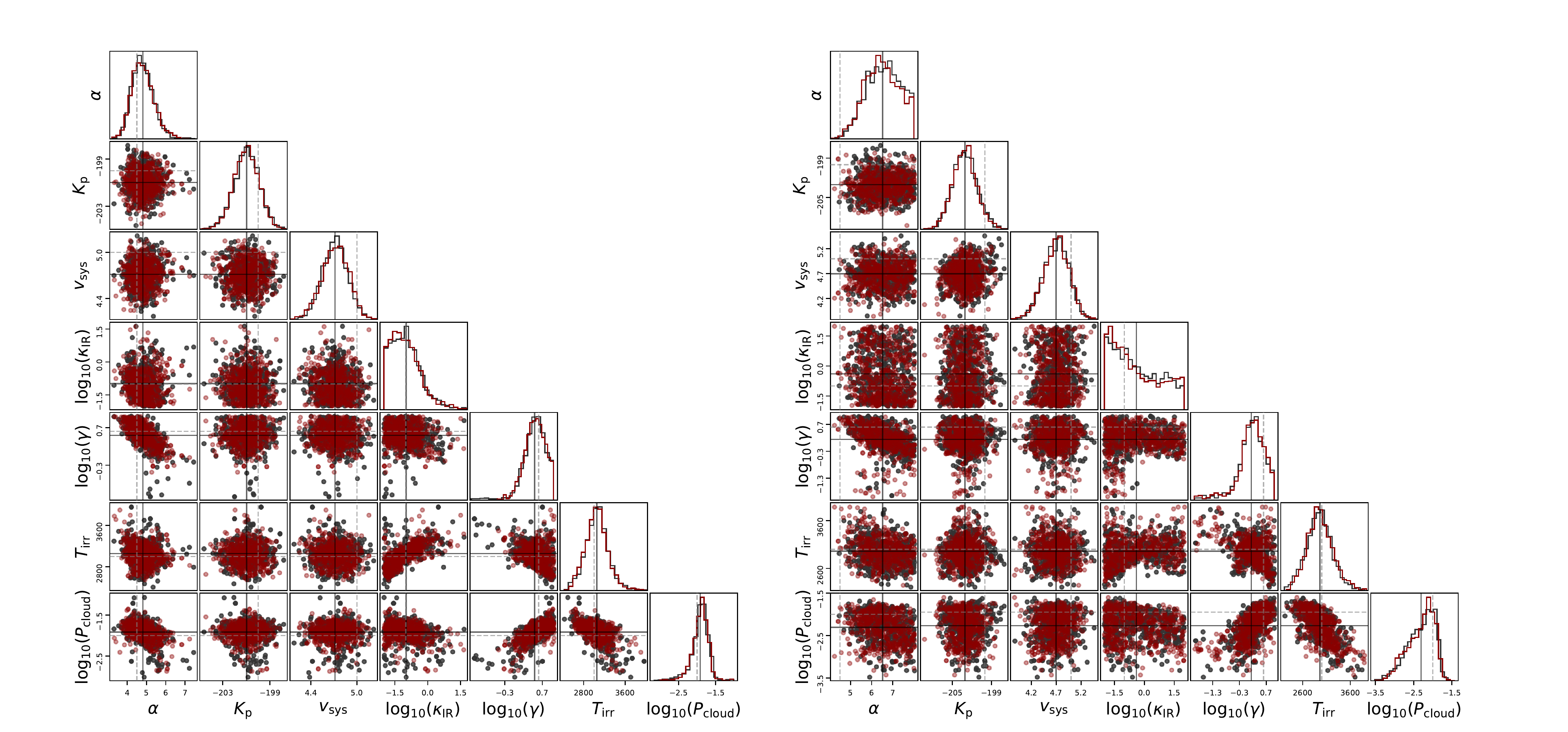}
\caption{Results of the first injection test using the alternative log likelihood formulations. The left plots show the 1D and 2D marginalised posterior distributions using the log likelihood with `nulled' $\beta$ but including the time- and wavelength-dependent noise (Eq.~\ref{eq:nulling}). The right plots show the same but with the uncertainties $\sigma$ also `nulled' (Eq.~\ref{eq:nulling-noiseless}). In this case the time-dependent noise is taken into account by partitioning the likelihood into separate spectra and summing; however, the wavelength dependence of the noise (within orders) is not accounted for.}
\label{fig:injection1_comparison}
\end{figure*}

We also performed retrievals on the same data using the `nulled' likelihoods given by Eqs.~\ref{eq:nulling} and \ref{eq:nulling-noiseless}. The resulting distributions from the MCMC are shown in Fig.~\ref{fig:injection1_comparison}. The nulled version that fully accounts for the time- and wavelength-dependent uncertainties provides a remarkably similar result to the `full' likelihood, with the one difference the omission of $\beta$, which is removed from the likelihood calculation. The version that does not account for the uncertainties also enables the recovery of parameters that are broadly consistent with the injected values; however, the resulting uncertainties are generally larger. This is unsurprising, as using optimal weights on the uncertainties will naturally result in better constraints. In this case, the time-dependence of the uncertainties is accounted for by partitioning the log likelihood into separate spectra (effectively optimising for a separate noise term for each spectrum), but this cannot account for the wavelength-dependent uncertainties. These are clearly important for UVES data, although we note that this may not be generally true. Nonetheless, we would recommend that the wavelength dependence of the noise should be accounted for where possible, either using the full likelihood calculation, or our generalised version of the \citet{2019AJ....157..114B} likelihood.

It is worth noting that the retrieved value for $\beta$ using our standard likelihood is lower than 1. At first glance this means that the uncertainties are overestimated by our simple estimation outlined in Sect.~\ref{sect:observations}. The underlying reason for this is that we used only a 5th-order PCA when estimating the noise, but have used 15 passes of {\sc SysRem} to remove systematics form the actual data. Algorithms such as PCA and {\sc SysRem} will inevitably remove some of the noise, and the added complexity of the {\sc SysRem} model applied to the data is the underlying reason why the scaling of the noise is found to be less than 1.  We note that we could generate the noise estimate using a 15th-order PCA to closely match the {\sc SysRem} model in which case $\beta\approx1$, but this is not expected to impact our results; the key goal of our noise estimate is simply to capture the time- and wavelength-dependent variation which is mostly captured by the first PCA component, and the $\beta$ parameter is optimised in our retrieval to ensure the correct scaling, as well as to marginalise over the uncertainty in the scaling. The suitability of our noise estimate is also shown by our injection and retrieval tests.
We finally note that this issue would still exist when using the `nulled' likelihoods. As noted these methods implicitly optimise for $\beta$ within each likelihood calculation even if they don't measure its value explicitly.

Next, we run a more complex injection and retrieval test using multiple atomic species, in this case with Fe, V and Mg, three species that were previously found in WASP-121b via high-resolution optical spectroscopy. The parameters of this injection test are again shown in Tab.~\ref{tab:injectiontests}. In this case, we also treat the (log) volume mixing ratios as free parameters. We also treat the convolution width $W_\mathrm{conv}$ as a free parameter, to test whether we can constrain the broadening of the spectral lines. This is generally something that is unknown in real datasets, and potentially provides information on the rotation and/or wind speeds in the planet's atmosphere. The results of this injection test are shown in Fig.~\ref{fig:injection2} and Tab.~\ref{tab:injectiontests}.

These results further demonstrate the effectiveness of our retrieval framework, and in addition show that we can constrain the convolution width using our log likelihood approach. In addition, this test shows that we can place constraints on the absolute abundances (roughly 0.5--0.7 on the log VMR for Fe and Cr). However, we note that these are highly degenerate with the scattering parameters (in this case $P_{\rm cloud}$), and limiting the cloud deck altitude informs the absolute abundance constraints (which may be a reasonable approach with physically motivated priors). Nonetheless, the 2D marginal distributions also show strong correlations between the abundances. This shows that we can constrain the abundance ratios, even if absolute abundances are poorly constrained due to other degeneracies. Given the nature of these degeneracies, we expect to be able to constrain relative abundances even where we have limited information on the scattering properties of the atmosphere. In the case of WASP-121b, the assumption of a well-mixed atmosphere is of greater concern. We discuss this issue later, but highlight that while there may be limitations with the atmospheric models, the underlying statistical and filtering framework perform well for a range of plausible scenarios, and show that we can constrain the scale of the model atmosphere, the broadening profile, the velocities, as well as marginalise over the noise properties (in this case a simple scaling term).

\subsection{Application to UVES data of WASP-121b}
\label{sect:application}

Now that we have validated our method on injected datasets, we turn our attention to the real signal of WASP-121b in our UVES data. In this case we pre-process the data as described in Sect.~\ref{sect:preprocessing} using 15 {\sc SysRem} iterations. We discard the first and last orders of the blue arm, as well as the first 21 orders and last order of the red arm. This is because the edge orders are not complete, and the red end of the red arm contains very low signal due to the dichroic. In addition, we also discard two further orders within the red arm, which fall on the gap between the detector mosaic (between the `lower' and `upper' red chips). We note that is no technical reason why we need to exclude low S/N data, as weighting by the uncertainties should optimally combine all the data. This was done purely to speed up the retrievals, where the likelihood calculation scales with the number of data points, and we can freely discard regions where we are confident there is no contribution to the signal.
We then perform a retrieval considering multiple species in the atmospheric model including Fe, Cr, Mg, V and Ti.
With the exception of Ti, these species have already been detected in WASP-121b \citep[with Fe, Cr, and V also found with our UVES dataset, see e.g.][]{2020MNRAS.493.2215G,2020A&A...641A.123H,2021MNRAS.506.3853M}. We include Ti in order to try and constrain the Ti/V ratio. In this case we fit for all of the T-P profile parameters, the abundances of the species, both scattering parameters, the convolution width, the planet velocities, and the scaling parameters. Tab.~\ref{tab:results} shows the assumed priors for each parameter. We perform separate fits for the blue and red arms, before also performing a combined fit. In the latter case, we allowed for independent values of $\beta$, velocities ($K_{\rm p}$ and $v_{\rm sys}$), and convolution width ($W_{\rm conv}$). In all cases we use the full likelihood (i.e. Eq.~\ref{eq:full}).

\begin{figure*}
\centering
\includegraphics[width=150mm]{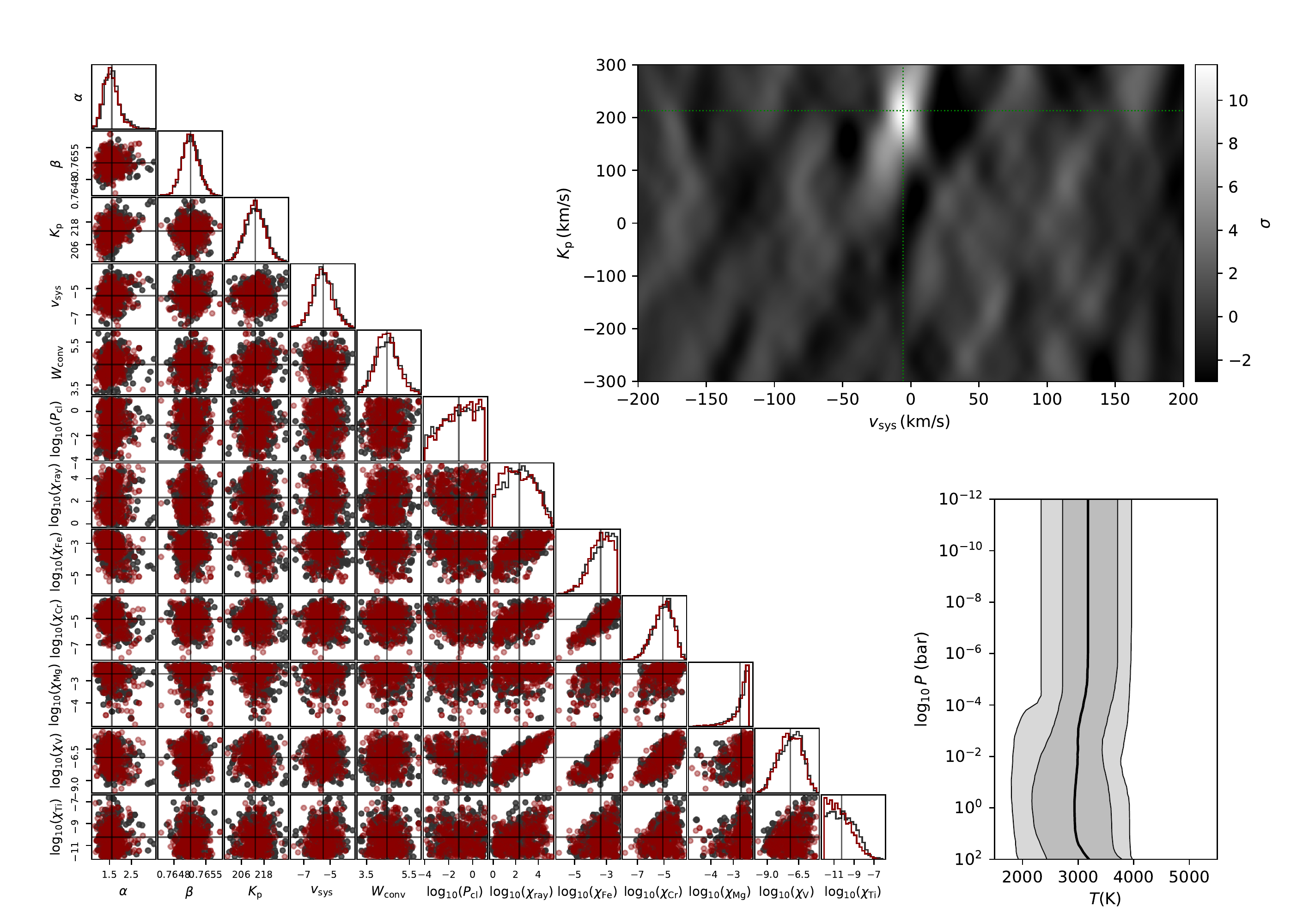}
\caption{Results of the retrieval for the blue data alone, using the same format as the injection tests. The solid lines show the means of the recovered parameter, and the dotted lines in the cross-correlation map shows the recovered values of $K_{\rm p}$ and $v_{\rm sys}$. We have removed the temperature profile parameters from the MCMC distributions to make room for the other parameters, but they are still free parameters in the MCMC.}
\label{fig:realdatablue}
\end{figure*}

\begin{figure*}
\centering
\includegraphics[width=160mm]{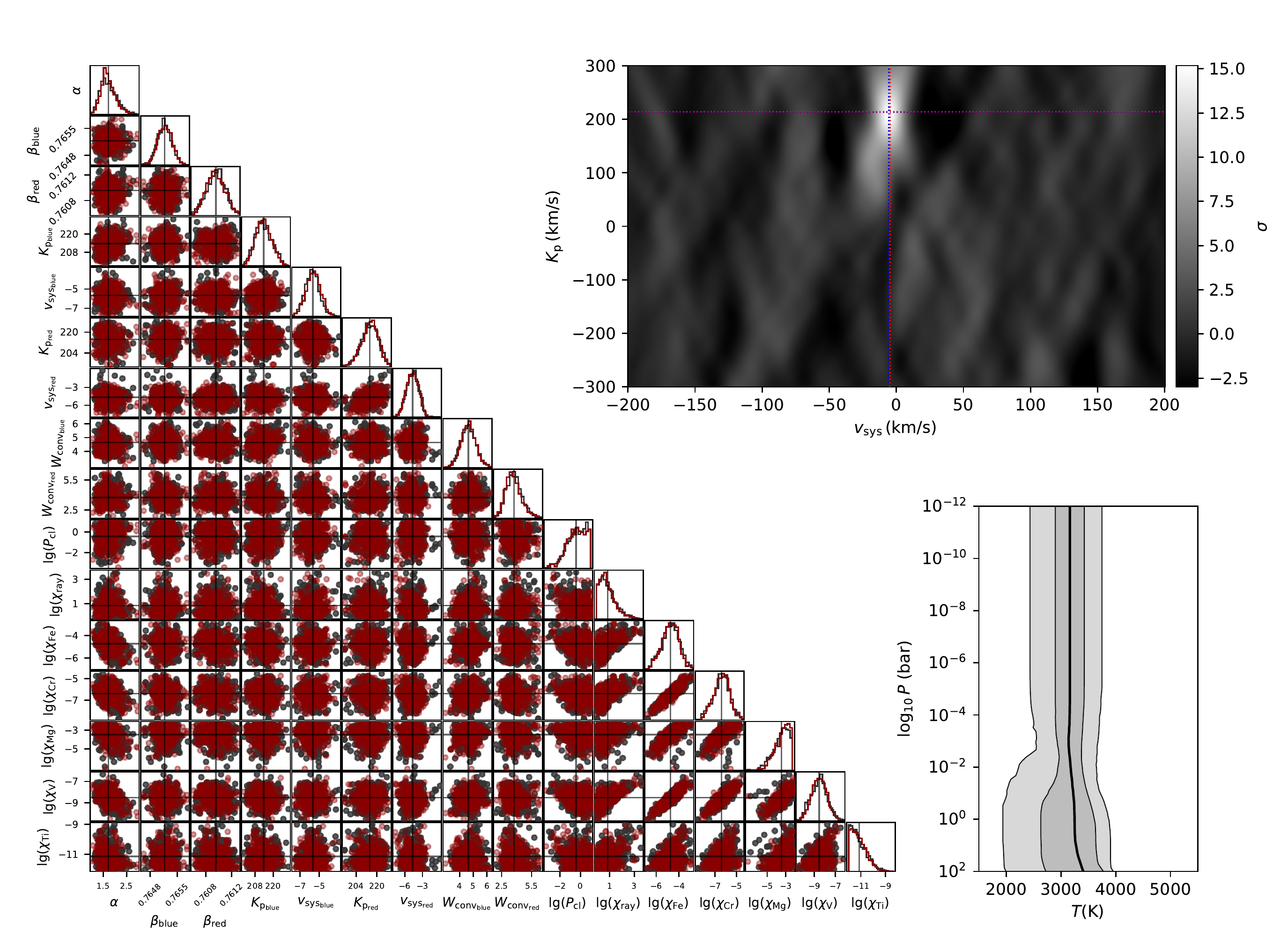}
\caption{Results of the MCMC for the combined red and blue datasets. In the cross-correlation map, the red and blue dotted lines show the recovered values of $K_{\rm p}$ and $v_{\rm sys}$ for each of the UVES arms, which are consistent. Otherwise the format is the same as for previous plots.}
\label{fig:realdataboth}
\end{figure*}

For each retrieval we run MCMC chains with a burn-in length of 200, chain length of 400, and 128 walkers.
The results for the blue arms and the combined fits are shown in Figs.~\ref{fig:realdatablue} and \ref{fig:realdataboth}. The equivalent fit for the red arm alone is shown in Fig.~\ref{fig:realdatared}. The results for all of the fits are provided in Tab.~\ref{tab:results}.
In general, the T-P profile parameters are not well constrained, and we get relatively weak constraints on the temperature in the upper atmosphere\footnote{Note that we removed the T-P parameters from the MCMC plots to make room for the other parameters, as they are shown in the T-P profile, and as noted are poorly constrained.}.
The velocities and scale parameters as well as the convolution width are all well constrained. The abundances of the species show complex correlations. As noted for the injection tests, this is natural as raising the cloud deck and raising all of the abundances results in a similar model -- a well known degeneracy in transmission spectroscopy \citep[see for example][we discuss the interpretation of this in the following section]{
2012ApJ...753..100B,2017MNRAS.470.2972H}. However, for each of the species that have been previously detected (Fe, Cr, Mg, V), we constrain their abundances to be non-zero in the combined fit (in the sense that they are inconsistent with the prior limits at the lower end). We therefore can consider these a detection from the MCMC distributions alone. All of these species are detected in the blue arm alone, and all with the exception of Mg are additionally found in the red arm (although Mg is preferentially shown to have high-abundance, pointing to a marginal detection).
Ti on the other hand is consistent with the lower prior limit, i.e. $\log_{10}(\chi_{\rm Ti}) = -12$, showing that we find no evidence for it in our data. Therefore we conclude that our data shows detections of Fe, Cr, Mg, V, and a non-detection of Ti.

\begin{figure*}
\centering
\includegraphics[width=180mm]{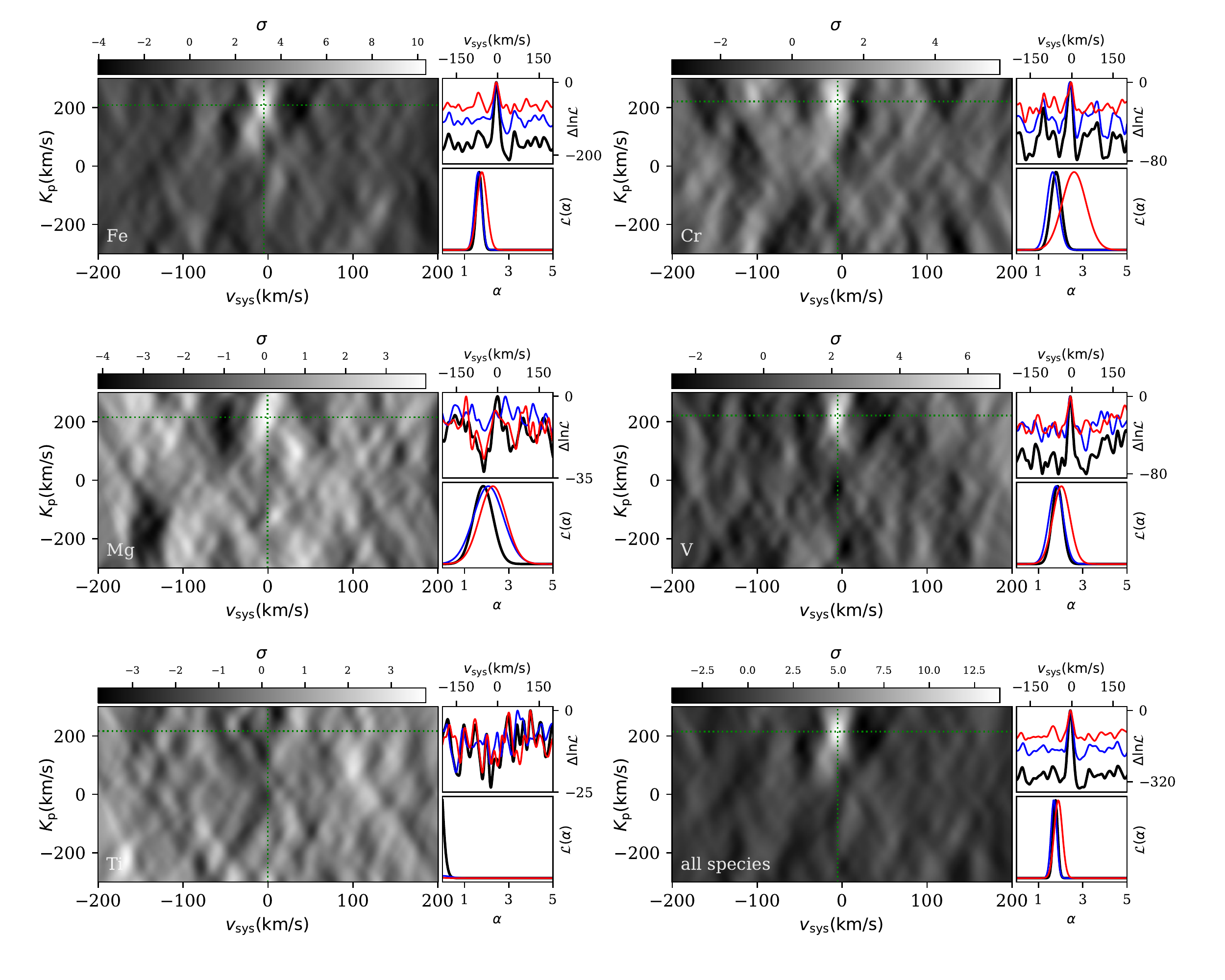}
\caption{Detections of the various species using `classical' cross-correlation analysis using models with each of the individual species, plus the combined model. The main plot shows the CCF map summed up over all orders, scaled to the standard deviation of an out-of-peak region of the map (i.e. standard `S/N' approach). The upper right plot for each species shows a slice through the maximum $K_{\rm p}$, converted to a log likelihood (shown after subtracting the maximum). 
The bottom right for each species shows the measured value of $\alpha$ for each species (conditioned on the optimum $v_{\rm sys}$ and $K_{\rm p}$), which is used to compute the detection significance.
The best-fit $K_{\rm p}$ and $v_{\rm sys}$ are marked by dotted lines, with the exception of Ti, where we use the velocities derived from the model with all species included. 
The blue, red and black lines correspond to the blue, red and both UVES arms, respectively. We detect Fe, Cr, Mg, and V, but not Ti, consistent with our retrieval analysis, and with previous studies of WASP-121b.}
\label{fig:detections}
\end{figure*}

To further demonstrate the detections (and non-detection of Ti), we also compute the CCFs for each species. We use the mean parameters from the MCMC distribution of the combined fits for the detected species\footnote{with the exception of the red arm of Mg, where we use the best-fit model, as the abundance is not well constrained}, and recompute the models with all but one of the species removed.
For Ti we use an enhanced value of $\log_{10}(\chi_{\rm Ti}) = -7$ to ensure sufficient lines are above the continuum when computing the CCF.
We then compute the CCF map for the blue and red arms combined. We also compute the likelihood from each of the CCF maps, using 100 samples of $\alpha$ ranging from 0 to 5, before determining the detection significance similarly to \citet{2020MNRAS.493.2215G}. This simply computes the conditional distribution of $\alpha$ (with all other parameters held at their optimal values), and extracts its mean and standard deviation. Finally we divide the mean by its standard deviation to get a detection significance. This method has the advantage that it does not rely on the CCF map at different values for $K_{\rm p}$ and $v_{\rm sys}$ to determine the underlying noise, which can in principle have different noise properties at the peak when compared to the `background' in the map.
The CCF maps, log likelihood slices, and conditional distribution of $\alpha$ (for optimal values of $K_{\rm p}$ and $v_{\rm sys}$)  are shown in Fig.~\ref{fig:detections}.
 We find detections of Fe, Cr, Mg and V at 12.6, 7.5, 4.1, and 7.5\,$\sigma$, respectively. For the combined model of all species, the atmospheric model is detected at 16.6\,$\sigma$. We do not get a peak at the `correct' value of Ti, indicating a non-detection. These (non-)detections are consistent with previous studies of the UVES datasets \citep{2021MNRAS.506.3853M}, as well as other studies of WASP-121b \citep[e.g.][]{2020A&A...641A.123H}. We note that in the case of Mg, \citet{2021MNRAS.506.3853M} do not find strong evidence for Mg from the cross-correlation analysis, although flagged it as a potentially interesting signal. Indeed, in the case of our CCF analysis alone, we would not conclude to have found a strong detection of Mg, as the CCF map contains significant negative peaks. Nonetheless, a positive CCF peak is found at consistent $K_{\rm p}$ and $v_{\rm sys}$ to the other species which supports its detection from the retrieval. In addition, the $\alpha$ value is constrained to be non-zero at more than 4\,$\sigma$, which further boosts our confidence in its detection, along with the retrieval analysis. Mg has also been detected in WASP-121b by \citet{2020A&A...641A.123H} and \citet{2021A&A...645A..24B}. Ultimately the models used in the CCF analysis with a single species are not the optimum way to detect them in exoplanet atmospheres, as this doesn't take into account other species that can mask a subset of lines, and/or change the effective line strengths of lines where they act as a pseudo-continuum.
 
\begin{table*}
\caption{Parameters recovered for the individual fits of the blue and red arms, as well as the combined fits, respectively.}
\label{tab:results}
\begin{tabular}{lllll}
\hline
\noalign{\smallskip}
Parameter & Prior & Value & Value & Value\\ 
~(arm)[units]  & ~ & (blue arm only) & (red arm only) & (combined)\\
\hline
\noalign{\smallskip}
~$\alpha$                               & $\mathcal{U}(0.1,6)$        & $1.57^{+0.37}_{-0.31}$           & $2.05^{+0.54}_{-0.46}$     & $1.68^{+0.37}_{-0.28}$        \\\noalign{\smallskip}
~$\beta(\rm blue)$                     & $\mathcal{U}(0.1,2)$        & $0.7652^{+0.0002}_{-0.0002}$           & -     & $0.7652^{+0.0002}_{-0.0002}$        \\\noalign{\smallskip}
~$\beta(\rm red)$                      & $\mathcal{U}(0.1,2)$        & -                                      & $0.7610^{+0.0001}_{-0.0001}$     & $0.7610^{+0.0002}_{-0.0001}$        \\\noalign{\smallskip}
~$K_{\rm p}(\rm blue)$ [km/s]                     & $\mathcal{U}(170,230)$      & $213.3^{+5.3}_{-5.2}$         & -      & $213.4^{+5.9}_{-5.2}$      \\\noalign{\smallskip}
~$v_{\rm sys} (\rm blue)$  [km/s]                  & $\mathcal{U}(-10,10)$       & $-5.57^{+0.78}_{-0.70}$          & -      & $-5.66^{+0.73}_{-0.75}$       \\\noalign{\smallskip}
~$K_{\rm p}(\rm red)$     [km/s]                 & $\mathcal{U}(170,230)$      & -                                      & $214.4^{+6.6}_{-7.2}$     & $214.9^{+6.1}_{-6.9}$      \\\noalign{\smallskip}
~$v_{\rm sys}(\rm red)$ [km/s]                   & $\mathcal{U}(-10,10)$       & -                                      & $-4.95^{+0.97}_{-0.92}$     & $-4.65^{+1.03}_{-1.06}$       \\\noalign{\smallskip}
~$W_{\rm conv} (\rm blue)$                  & $\mathcal{U}(1,50)$         & $4.44^{+0.55}_{-0.51}$           & -      & $4.64^{+0.55}_{-0.55}$        \\\noalign{\smallskip}
~$W_{\rm conv} (\rm red)$                   & $\mathcal{U}(1,50)$         & -                                      & $4.15^{+0.95}_{-0.87}$     & $3.68^{+0.83}_{-0.72}$        \\\noalign{\smallskip}
~$\log_{10}(\kappa_{\rm IR})$ [m$^2$/kg]                & $\mathcal{U}(-4,0)$         & $-2.35^{+1.02}_{-1.07}$          & $-2.14^{+0.91}_{-1.21}$     & $-2.54^{+1.12}_{-0.97}$       \\\noalign{\smallskip}
~$\log_{10}(\gamma)$                          & $\mathcal{U}(-2,2)$         & $0.14^{+0.64}_{-0.51}$           & $0.44^{+0.52}_{-0.55}$     & $-0.07^{+0.53}_{-0.28}$       \\\noalign{\smallskip}
~$T_{\rm irr}$   [K]                       & $\mathcal{U}(2000,4500)$  & $2960^{+420}_{-490}$    & $2990^{+430}_{-540}$           & $3180^{+240}_{-370}$ \\\noalign{\smallskip}
~$\log_{10}(P_{\rm cl})$                      & $\mathcal{U}(-4,1)$         & $-1.04^{+1.41}_{-1.67}$          & $-2.73^{+1.31}_{-0.83}$     & $-0.33^{+0.93}_{-1.03}$       \\\noalign{\smallskip}
~$\log_{10}(\chi_{\rm ray})$                  & $\mathcal{U}(-0.068,10)$    & $2.29^{+1.50}_{-1.45}$           & $1.02^{+0.99}_{-0.79}$   &   $0.72^{+0.78}_{-0.51}$     \\\noalign{\smallskip}
~$\log_{10}(\chi_{\rm Fe})$                    & $\mathcal{U}(-12,-2.3)$     & $-3.28^{+0.66}_{-0.80}$          & $-4.11^{+0.84}_{-0.78}$  &  $-4.72^{+0.71}_{-0.79}$    \\\noalign{\smallskip}
~$\log_{10}(\chi_{\rm Cr})$                    & $\mathcal{U}(-12,-2.3)$     & $-5.01^{+0.70}_{-0.85}$          & $-5.57^{+0.91}_{-0.86}$  &  $-6.36^{+0.69}_{-0.81}$    \\\noalign{\smallskip}
~$\log_{10}(\chi_{\rm Mg})$                    & $\mathcal{U}(-12,-2.3)$     & $-2.56^{+0.19}_{-0.47}$          & $-3.49^{+0.90}_{-6.10}$  &  $-3.33^{+0.67}_{-0.99}$    \\\noalign{\smallskip}
~$\log_{10}(\chi_{\rm V})$                     & $\mathcal{U}(-12,-2.3)$     & $-7.11^{+0.87}_{-1.02}$          & $-7.74^{+0.82}_{-0.91}$  &  $-8.51^{+0.72}_{-0.77}$    \\\noalign{\smallskip}
~$\log_{10}(\chi_{\rm Ti})$                    & $\mathcal{U}(-12,-2.3)$     & $-10.38^{+1.40}_{-1.12}$         & $-10.67^{+1.03}_{-0.94}$ &  $-11.26^{+0.78}_{-0.52}$      \\\noalign{\smallskip}
\hline
\noalign{\smallskip}
\end{tabular}
\end{table*}

\begin{figure}
\centering
\includegraphics[width=80mm]{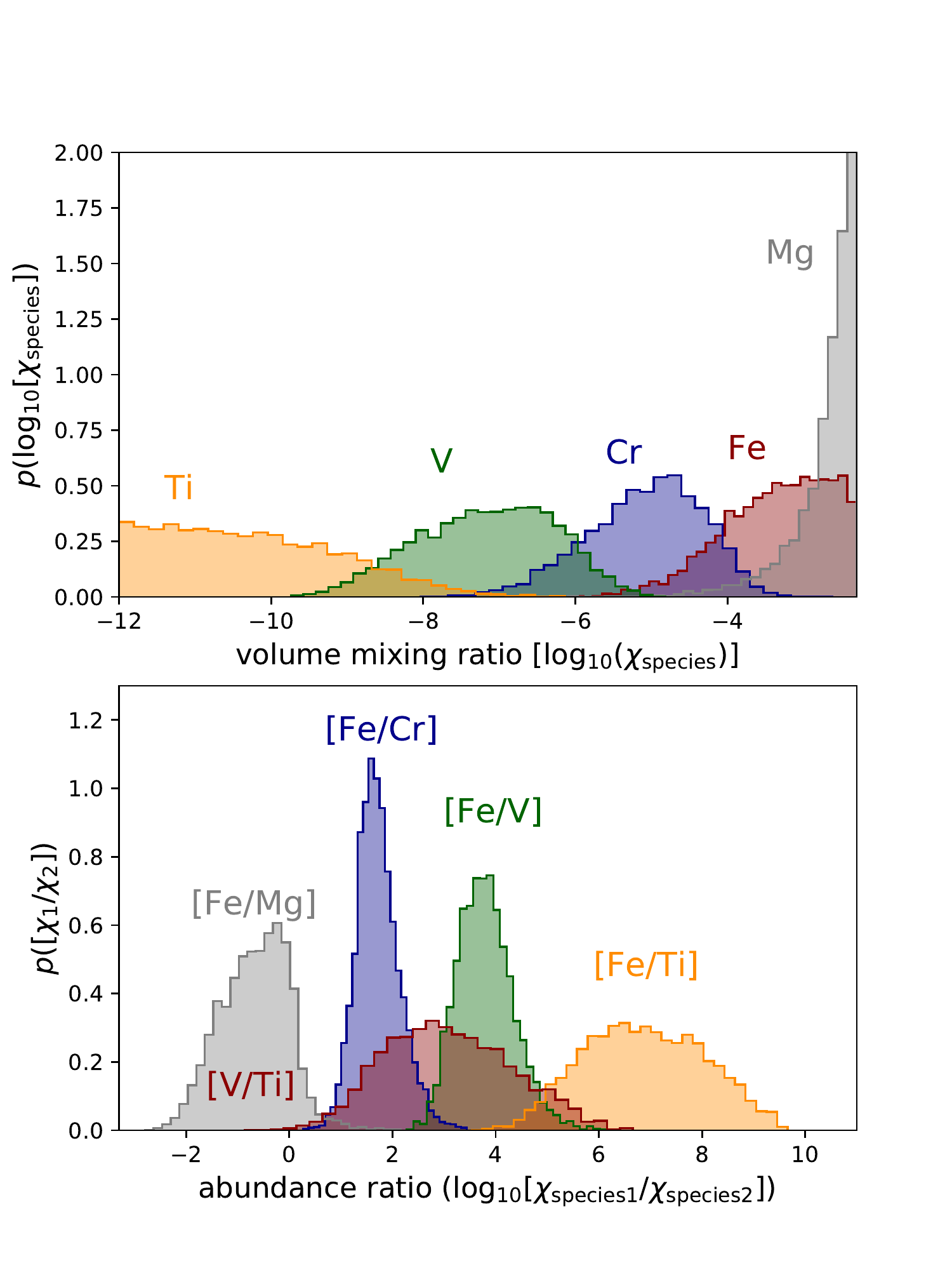}
\caption{Abundances and relative abundances for WASP-121b from the retrieval performed on the blue UVES arm}
\label{fig:realdatablue_abundances}
\end{figure}

\begin{figure}
\centering
\includegraphics[width=80mm]{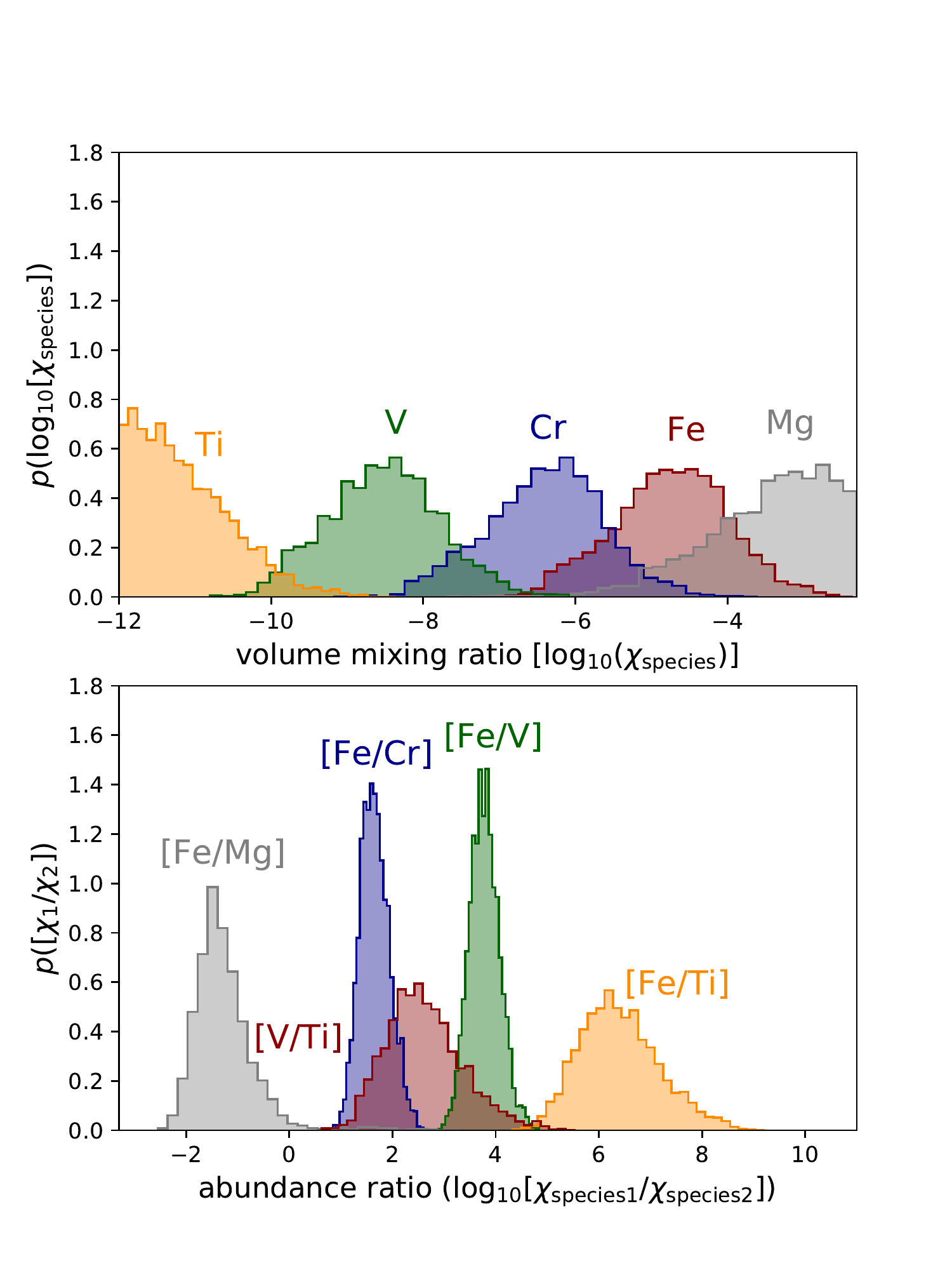}
\caption{Abundances and relative abundances for WASP-121b from the retrieval performed on both UVES arms}
\label{fig:realdataboth_abundances}
\end{figure}

\section{Discussion}
\label{sect:discussion}

Our analyses of the UVES data of WASP-121b clearly reveals detections of Fe, Cr, Mg and V, with detection significances of 12.6, 7.5, 4.1, and 7.5\,$\sigma$, respectively. These detections are consistent with previous analysis of this planet and indeed this dataset \citep[e.g.][]{2020MNRAS.493.2215G,2020A&A...641A.123H,2021MNRAS.506.3853M}. The species are also detected at consistent velocities with previous studies, showing a typical net blueshift of $\sim$5km/s.
Beyond simply detecting the individual species, our retrieval framework also allows us to constrain the absolute abundances of each species, the temperature-pressure profile, velocity broadening, the vertical extent of the atmosphere, as well as obtain robust uncertainties on the velocity amplitude and offset of the planet, 
all while marginalising over the noise properties of the data (in this case a simple scaling term, but this can easily be generalised). This is made possible by computing the likelihood of the model fit to the data rather than using the cross-correlation function, as well as using a filtering technique to account for the effects of {\sc SysRem} on the exoplanet's signal. 

The constraints on the temperature-pressure profile indicate a temperature of $\approx$3160$\pm$290\,K in the upper atmosphere, consistent with previous results from this dataset \citep{2020MNRAS.493.2215G}. The temperature is poorly constrained deeper in the atmosphere, as expected. The constraints are not strong enough to say anything meaningful about the existence or non-existence of a temperature inversion, and this is again consistent with our expectations, where we would expect emission spectra to be more sensitive to the temperature profile of the atmosphere. 
Our retrieved temperature is similar to previous retrievals of the dayside temperature, and significantly higher than temperatures retrieved for the terminator \citep[as well as upper limits on the nightside, e.g.][]{2018AJ....156..283E,2020A&A...637A..36B,2021MNRAS.503.4787W}. This is naturally explained by the fact that at high resolution we are probing higher up in the atmosphere, and coupled with the temperature inversion, means that we should expect higher temperatures in the upper atmosphere.

The abundance constraints are in general relatively broad, with uncertainties typically of order 1\,dex, and in some cases are constrained by the prior distributions and by the assumed scattering properties (and in turn their priors).
Nonetheless, the strong correlations between the absolute abundances of some of the species indicates that the {\it relative} abundances of the species can be constrained much more precisely. 
This is consistent with our expectations from transmission spectroscopy, where there are strong degeneracies between the cloud properties in the atmosphere and the absolute abundances \citep[see e.g.][]{2012ApJ...753..100B,2017MNRAS.470.2972H}, but relative abundances can be constrained from the relative line strengths of different species (see \citealt{2018arXiv180407357S} for an overview of the information content of transmission spectra, and \citealt{2015MNRAS.446.2428S} for an application to real data). While for ground-based high-resolution observations we filter out the planet's continuum, we are still sensitive to line strengths above the continuum over a broad wavelength range, allowing us to constrain relative abundances so long as we marginalise over effects that can alter these relative line strengths. This includes the scattering properties and temperature structure.
In addition, we must also account for the affects of the filtering on the model atmosphere which also distorts line strengths, as discussed at length in Sect.~\ref{sect:preprocessing_model}.

Fig.~\ref{fig:realdatablue_abundances} shows a closer look at the abundances of the five species from the blue arm. As noted, all species except Ti are inconsistent with the lower limit of the prior, showing that they are detected in our data. However, the abundance distributions are non-zero over several orders of magnitude. The lower panel of Fig.~\ref{fig:realdatablue_abundances} instead shows the {\it relative} abundance constraints on the same scale, where it is clear that the constraints are much tighter.
We find $\log_{10}(\chi_{\rm Fe}/\chi_{\rm Cr})$\,=\,1.66$\pm$0.28, $\log_{10}(\chi_{\rm Fe}/\chi_{\rm V})$\,=\,3.78$\pm$0.29 and $\log_{10}(\chi_{\rm Fe}/\chi_{\rm Mg})$\,=\,-1.26$\pm$0.60. 
We also find $\log_{10}(\chi_{\rm V}/\chi_{\rm Ti})$\,=\,2.62$\pm$0.74, but note that the upper limit is truncated by the lower limit of the Ti abundance prior and therefore we interpret this as a lower (1\,$\sigma$) limit of $\log_{10}(\chi_{\rm V}/\chi_{\rm Ti})$\,$\geq$1.88.
It is also important to note that the Fe and Mg abundances are consistent with the upper prior limit, which roughly corresponds to all of the gas {\it except} H and He being contained within that species. While this may appear unphysical, it is simply a result of the known degeneracies and limitations in our forward model, so we choose not to over-interpret the absolute abundances. For example, we do not adjust the mean molecular weight in our model in response to the increased abundances of heavy elements, which would be a potential way to constrain this effect (coupled with suitable constraints on $\alpha$); however, this would also require us to consider the total metallicity of the atmosphere, rather than only the species directly detected. This is currently beyond the capabilities of our atmospheric model, which we developed to be as simple and as fast as possible in order to enable high-resolution retrievals.
It is also worth noting that the exosphere of WASP-121b has been directly detected, showing evidence of Fe\,{\sc ii} and Mg\,{\sc ii} \citep{2019AJ....158...91S} extending up to several planetary radii. As we have assumed an atmosphere in hydrostatic equilibrium, it is possible that unphysically large abundances could be an indication that the species extends over more scale heights than predicted in our model, and therefore a signature of atmospheric escape, or at least consistent with it. In addition, \citet{2020A&A...641A.123H} found that the Mg signal required much larger line strengths than the other species, consistent with our expectation from an escaping atmosphere.

A natural question to ask is whether including a species that doesn't follow the hydrostatic atmosphere can modify our inference of the other species. Fig.~\ref{fig:realdatablue_nomg_abundances} shows the results from an identical retrieval, except leaving Mg out of the model. It is clear that while this has a minor impact on the absolute abundances, the relative abundances of the remaining species remain consistent, so we can be reasonably confident that the inclusion of Mg does not impact the relative abundances of the other species. However, the assumption of a well-mixed atmosphere, as well as the chemistry of the atmosphere (e.g. we know that some of the species are also ionised in the upper atmosphere or perhaps trapped within molecules), will of course impact measurement of relative abundances of each of the species, as well as their interpretation as elemental abundances. We highlight that this is a limitation of our forward model atmosphere, and not of the statistical framework. One possible solution from within the statistical framework might be to use independent values of $\alpha$ to allow for different species to extend over differing numbers of scale heights; however while technically feasible, a simple implementation of this would be highly degenerate with the abundances. A better solution would be to incorporate the physics of an escaping atmosphere into the atmospheric model (which could alternatively be folded into a species-dependent scale `function', rather than scale parameter $\alpha$), as well as the chemical profiles in the atmosphere. However, this is beyond the capabilities of our current model, and we therefore leave this for future work.
One further caveat is that we use a simplified scattering model and ignore other potentially important continuum opacities (e.g. H$^-$). Therefore the absolute values of the cloud deck and Rayleigh factor should not be over-interpreted. In addition, it is possible that these parameters are impacted by other opacity sources not accounted for in our model, for example atomic or molecular features with strong optical absorption could drive up the Rayleigh `abundance'. However, our expectation is that as long as the scattering model is flexible enough to capture the broad continuum, the relative abundances can be constrained. {\it We highlight that it is important to take all of these limitations into account when interpreting our results.}

Nonetheless, taken at face value, the relative abundances of Fe, Cr and V are consistent with solar abundances \citep[e.g.][]{2009ARA&A..47..481A}\footnote{Solar values are $\log_{10}(\chi_{\rm Fe}/\chi_{\rm Cr})$\,=\,1.86, $\log_{10}(\chi_{\rm Fe}/\chi_{\rm V})$\,=\,3.57, $\log_{10}(\chi_{\rm Fe}/\chi_{\rm Mg})$\,=\,-0.10, and $\log_{10}(\chi_{\rm Fe}/\chi_{\rm Ti})$\,=\,2.55.}.
The measured Fe/Mg ratio is sub-solar, presumably due to the overestimate of $\chi_{\rm Mg}$ resulting from the non-hydrostatic atmosphere noted earlier that our model does not account for.
Furthermore, Ti is significantly less abundant than V, where it would be expected to be roughly an order of magnitude larger if both species were purely in atomic form\footnote{Solar value of $\log_{10}(\chi_{\rm Ti}/\chi_{\rm V})$\,=\,1.02}, although chemical models of ultra-hot Jupiters typically predict that at least some of these species should be trapped in TiO and VO \citep[e.g.][]{
2008ApJ...678.1419F}. However, despite being strong absorbers at optical wavelengths, TiO and VO have not yet been detected in WASP-121b \citep{2020A&A...636A.117M}. Our constraints on Ti/V are consistent with previous findings from \citet{2020A&A...641A.123H}. Coupled with the non-detection of TiO, they concluded that Ti may be depleted via a cold-trap mechanism. Our models did not consider TiO and VO, due to their non-detection; however, it is important to note that limitations in the accuracy of their line lists, particularly VO, may have prohibited detection at high resolution \citep{2020A&A...636A.117M}. Indeed, at low-resolution, there is evidence for the presence of VO \citep{2018AJ....156..283E}, although a high-resolution detection will be required to unambiguously confirm its presence, particularly given that the optical spectrum of WASP-121b is complicated by the present of many atoms and ions with strong (and complex) optical spectra. The interpretation is further complicated by the possible signs of variability in WASP-121b's atmosphere \citep{2021MNRAS.503.4787W}, which makes comparisons across datasets even trickier. 

Ignoring these complications in interpreting the abundances, as noted, the Fe, Cr and V abundance ratios are consistent with solar values. This suggests that there is no evidence for fractionisation of these species from the material that ultimately formed WASP-121b. Unfortunately as all of these species are refractory, with similar condensation temperatures, they are not an ideal selection of species to trace the formation history of the planet \citep{2021ApJ...914...12L}. Nonetheless, the fact that they are constrained to roughly 0.3\,dex is extremely encouraging for future measurements of relative abundances from high-resolution transmission spectroscopy in efforts to constrain refractory-to-volatile elemental ratios. Similar techniques at near-infrared wavelengths targeting C and O bearing species should be able to constrain C/O ratios in a similar way, again leading to potential constraints of planet formation pathways \citep{2011ApJ...743L..16O}. The combination of optical and near-infrared observations should be able to target both refractory and volatile species.

We also constrain the scale factor, $\alpha$, to be $1.68^{+0.37}_{-0.28}$ in the combined fits, and find consistent values from the individual analyses of the red and blue arms. This shows that the amplitude of the model is slightly larger than predicted from our 1D forward model. We consider it important to marginalise over the scale factor $\alpha$, as our simplified model atmosphere does not account for all possible effects that might increase the model amplitude, or equivalently the (vertical) extent of the atmosphere. For example, we do not take into account the fact that at high temperatures, molecular hydrogen dissociates into atomic hydrogen, therefore reducing the mean molecular weight and inflating the local atmospheric scale height\footnote{To first order the mean molecular weight is is reduced from 2.33 to $\sim\,$1.3 (ignoring free electrons) for complete dissociation of hydrogen, therefore the scale height would increase by a factor of $\sim$\,1.8.}. This is a likely explanation for the enhanced value of $\alpha$ in the case of WASP-121b.
Another possibility is that the exospheric escape also impacts the species in our retrieval, effectively stretching the atmosphere over more scale heights than predicted from a hydrostatic atmosphere.
If we did not include the scale factor as a free parameter, there would be tension between the temperature constraints from the relative line strengths vs the vertical structure of the atmosphere, and this may lead to biases in the retrieval. In this case, including the scale factor means that the determination of the atmospheric temperature-pressure profile is purely down to the relative line strengths of the species. Setting $\alpha=1$, and incorporating all such effects into the model atmosphere would likely enable more stringent constraints on the T-P profile and perhaps abundances. However, as noted this is beyond the capabilities of our current atmospheric model. 

Finally we also obtain velocity and broadening constraints from our retrievals. The velocities determined from the red and blue arms are consistent, as are the widths of the broadening kernels. We find a small blueshift (relative to the stellar rest frame) of $-5.7^{+0.7}_{-0.8}$\,km/s and $-5.0^{+1.0}_{-0.9}$\,km/s for the blue and red arms, respectively, consistent with previous measurements of species in WASP-121b \citep[e.g.][]{2020A&A...635A.205B,2020MNRAS.493.2215G,2020A&A...641A.123H,2020MNRAS.494..363C}, enabling a direct measurement of atmospheric dynamics in the system. The blueshift is a result of a net shift of material from dayside to nightside at the pressures probed, or some asymmetry in the absorption around the limbs, potentially due to different continuum absorption in the leading and trailing limbs of the planet, or changing chemistry due to differing temperatures \citep[e.g.][]{
2020Natur.580..597E}. The consistency of the blueshifts for different species more likely points to one of the former explanations. The existence of an escaping exosphere on WASP-121b may complicate this picture even further.

We also measure a broadening width of 4.64$\pm$0.55 and $3.68^{+0.83}_{-0.72}$ for the blue and red arms, respectively. These are in units of standard deviations of a Gaussian kernel at a resolution of R=200,000, or velocity of $\approx$1.5\,km/s. Therefore the broadening kernel width transforms into FWHMs of 16.4$\pm$1.9\,km/s and $13.0^{+2.9}_{-2.5}$\,km/s in the blue and red arms, respectively. The rotational velocity of WASP-121b, assuming it is tidally locked, is 7.4\,km/s at its equator (i.e. $\approx$15\,km/s peak-to-peak). Therefore our measurements of the broadening do not indicate clear evidence of broadening beyond rotation, e.g. due to equatorial winds. However, it is possible that the species are not uniformly distributed around the limb of the planet, further complicating the interpretation. In addition we need to take into account a more realistic broadening kernel associated with the transit geometry. Nonetheless, our work is a proof-of-concept that broadening parameters can be constrained from high-resolution retrievals, and more complex kernels with multiple parameters will be explored in future work.

\section{Conclusions}
\label{sect:conclusion}

We have presented an extension of our high-resolution retrieval framework previously introduced in \citet{2020MNRAS.493.2215G}, by developing a new filtering technique that can mimic the effects of {\sc SysRem} or PCA on the exoplanet's signal when removing stellar and telluric lines from the data. This is of key importance for high-resolution retrievals, where the filtering of the stellar and telluric signals will also distort the underlying exoplanet signal, and therefore the forward model atmosphere must also be modified in the same way. Crucially, this approach is fast enough so that it can be applied within each likelihood calculation without becoming the major bottleneck in the retrieval\footnote{For example, our final retrieval on the blue arm, with $\approx$9 million data points and 76,800 posterior samples, takes $\approx$\,2hours on a quad-core laptop (although note that a substantial fraction of steps are outside the prior range and therefore the likelihood need not be computed; target acceptance rate is $\approx$25\,\%). The bottleneck in each calculation is primarily the interpolation of the 1D model to the 3D data array (i.e. order $\times$ time $\times$ wavelength)}.
We use a series of injection tests to show that we can recover robust constraints on a wide range of model parameters using this framework. 

We also apply our new framework to perform a re-analysis of high-resolution UVES data of a transit of the ultra-hot Jupiter WASP-121b, previously analysed in \citet{2020MNRAS.493.2215G}. We detect Fe, Cr, V and Mg in the atmosphere with significances of 12.6, 7.5, 4.1, and 7.5\,$\sigma$, respectively, and we do not find evidence for Ti. These results are consistent with previous analyses of WASP-121b. Our retrieval framework also allows us to constrain the abundances, temperature-pressure profile, and broadening parameters, while simultaneously marginalising over the noise properties of the dataset. The temperature-pressure profile is constrained to $\approx$3160$\pm$290\,K in the upper atmosphere, consistent with previous results. The absolute abundances of the detected species are typically constrained to $\sim$1\,dex. Several combinations of species also show strong correlations and therefore the {\it relative} abundances can be constrained more precisely. We find $\log_{10}(\chi_{\rm Fe}/\chi_{\rm Cr})$\,=\,1.66$\pm$0.28, $\log_{10}(\chi_{\rm Fe}/\chi_{\rm V})$\,=\,3.78$\pm$0.29 and $\log_{10}(\chi_{\rm Fe}/\chi_{\rm Mg})$\,=\,-1.26$\pm$0.60. Importantly, these constraints depend on the assumption of a well-mixed atmosphere, and the latter in particular is likely to be impacted by the escaping atmosphere which is not taken into account in our model. If taken at face value, the relative abundances of Fe, Cr and V are all consistent with solar values, indicating that they are all accreted at similar ratios from the proto-planetary disk. Given that these are all refractory species, these combinations of relative abundances do not reveal much about the planet's formation history \citep{2021ApJ...914...12L}. Nonetheless, the precision by which the relative abundances can be constrained is encouraging for future constraints of refractory-to-volatile species, as well other informative abundance ratios such as C/O, which could be constrained from near-infrared transmission spectroscopy. The combination of optical and near-infrared high-resolution transmission spectroscopy datasets, which is (in principle) straightforward to perform, should provide interesting constraints on refractory-to-volatile elemental ratios for ultra-hot Jupiters.
Furthermore, these techniques, including the fast-filtering approach, can also be applied to high-resolution emission and phase curve observations to provide further constraints on atmospheric parameters.

Finally, this work demonstrates that we can also constrain additional dynamical information such as the (mean) broadening of the spectral features, as well as the net blue- or red-shift of the model atmosphere. More realistic broadening kernels, and independent kernels for each species should allow us in principle to probe the dynamics in more detail. Our retrieval framework, along with other complementary approaches such as \citet{2019AJ....157..114B}, show that high-resolution time-series observations are not limited to simply detecting atmospheric features, but can also constrain a rich array of model parameters including (relative) abundances, scattering properties and constraints on the temperature-pressure profiles, previously thought to require `traditional' low-resolution observations (i.e. where the continuum is not filtered from the data). Additionally, high-resolution observations have the added advantage that they can be harnessed to constrain the atmospheric dynamics. We have shown that our statistical framework is able to robustly perform retrievals on high-resolution observations, fully taking into account the filtering of the model atmosphere in an efficient manner. The only remaining limitation is coupling this statistical framework to sufficiently complex atmospheric models at high-resolution to fully exploit the power of the high-resolution Doppler-resolved spectroscopy.

\section*{Acknowledgements}

We are extremely grateful to the anonymous referee for careful reading of the manuscript and comments that improved the clarity of the paper.
This work is based on observations collected at the European Organisation for Astronomical Research in the Southern Hemisphere under ESO programme 098.C-0547. N. P. G. gratefully acknowledges support from Science Foundation Ireland and the Royal Society in the form of a University Research Fellowship.
We are grateful to the developers of the {\sc NumPy, SciPy, Matplotlib, iPython} and {\sc Astropy} packages, which were used extensively in this work \citep{Jones_2001,
2007CSE.....9...90H,2007CSE.....9c..21P,2013A&A...558A..33A}.
We are also grateful to the developers of petitRADTRANS for making their code available, which was invaluable in testing our atmospheric model.

\section*{Data Availability}

The observations underlying this analysis are publicly available in the ESO Science Archive Facility (\url{http://archive.eso.org}) under program name 098.C-0547. Data products will be shared on reasonable request to the corresponding author.

%




\bibliography{export-bibtex.bib} 
\bibliographystyle{mnras} 



\appendix
\section{Extra material}
%

Here we include additional plots of posterior distributions from our MCMC fits of the injected and real datasets. See main text in Sects.~\ref{sect:injections} and \ref{sect:application} for details.

\begin{figure*}
\centering
\includegraphics[width=150mm]{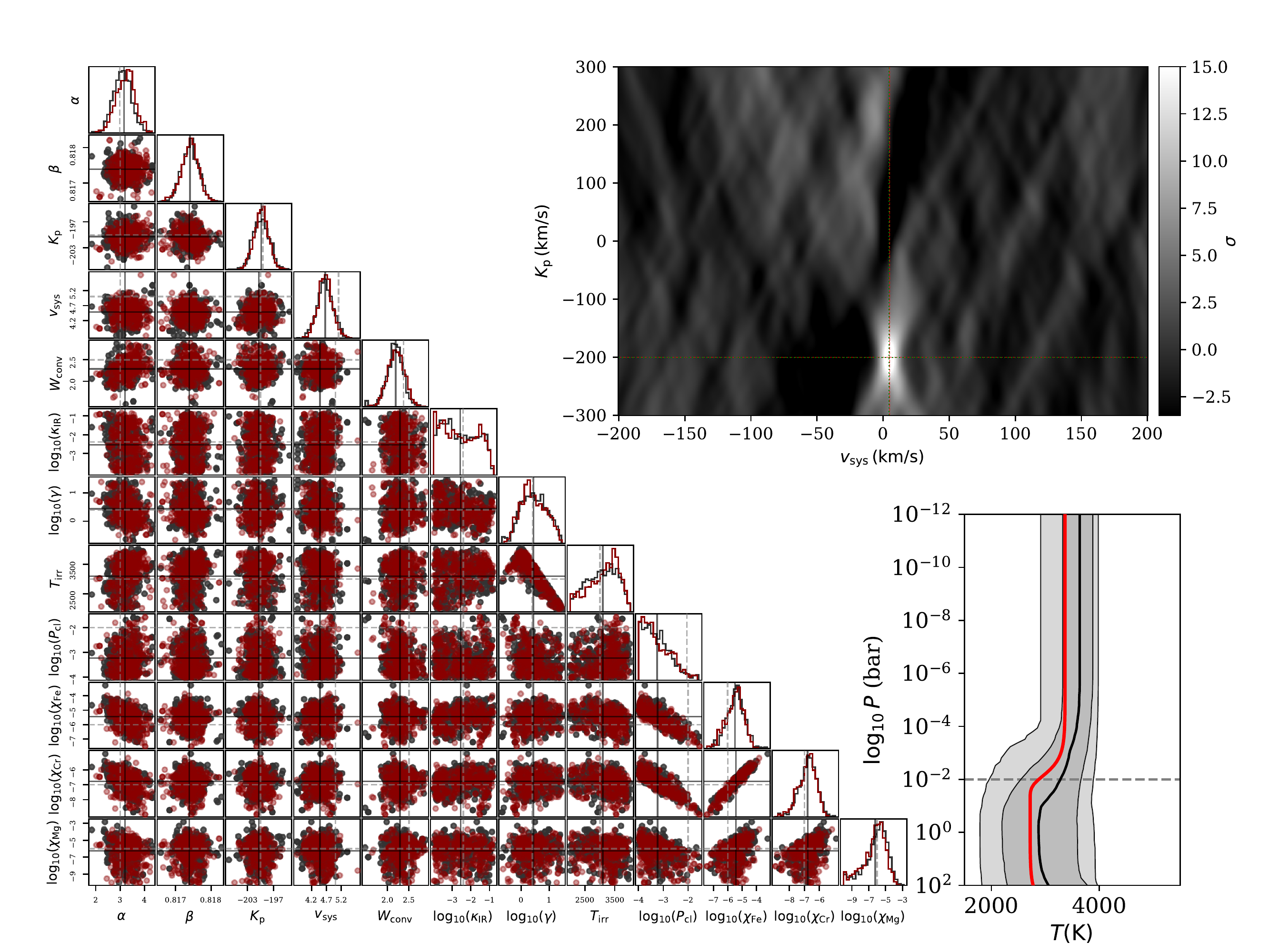}
\caption{Results of the second injection test, using the same formatting as Fig.~\ref{fig:injection1}. Compared to the first example, we now inject multiple species, and treat the convolution width and abundances as free parameters.}
\label{fig:injection2}
\end{figure*}

\begin{figure*}
\centering
\includegraphics[width=140mm]{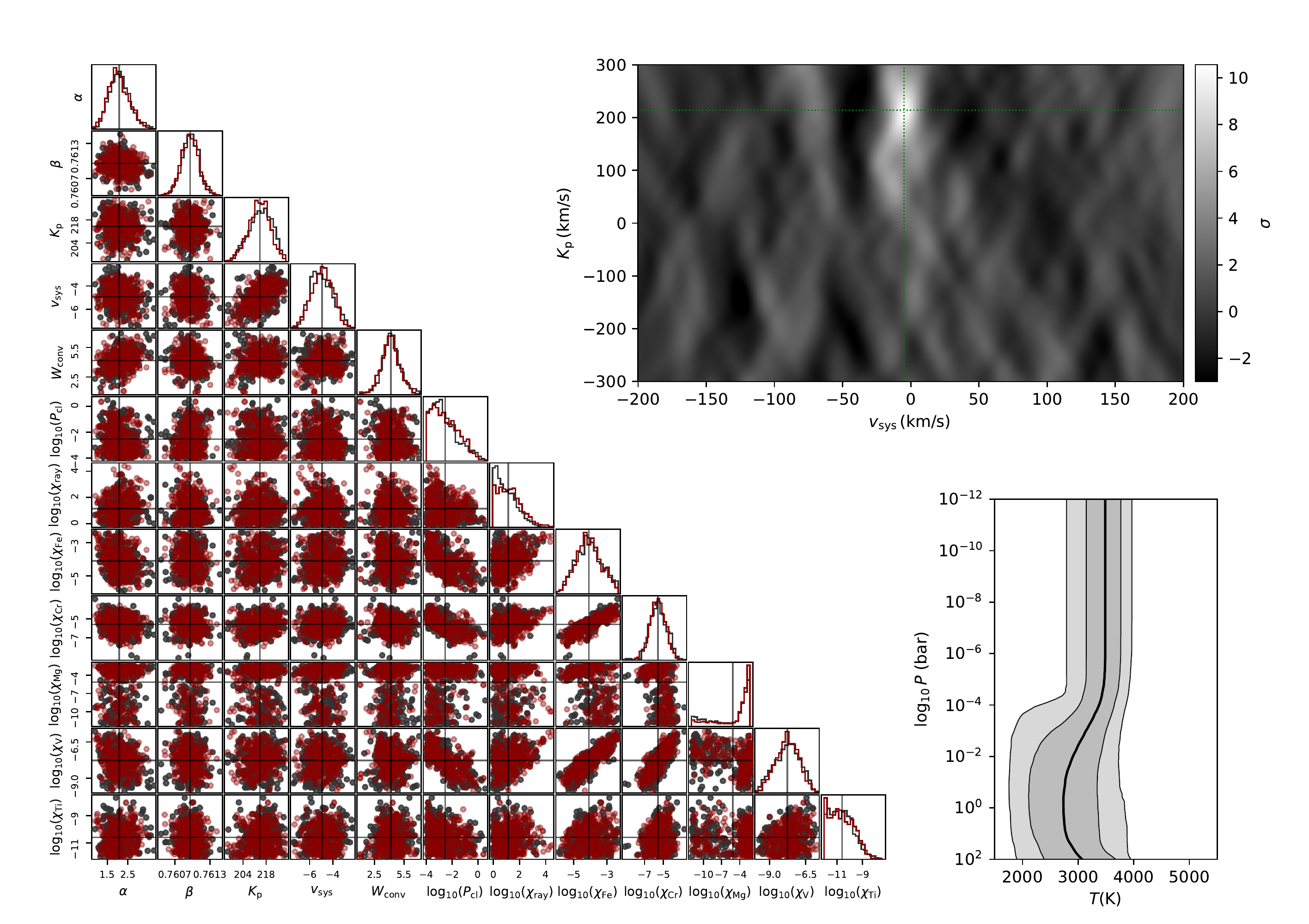}
\caption{Results of the retrieval of the UVES data of WASP-121b for the red data alone. Formatting is the same as for previous plots.}
\label{fig:realdatared}
\end{figure*}

\begin{figure}
\centering
\includegraphics[width=70mm]{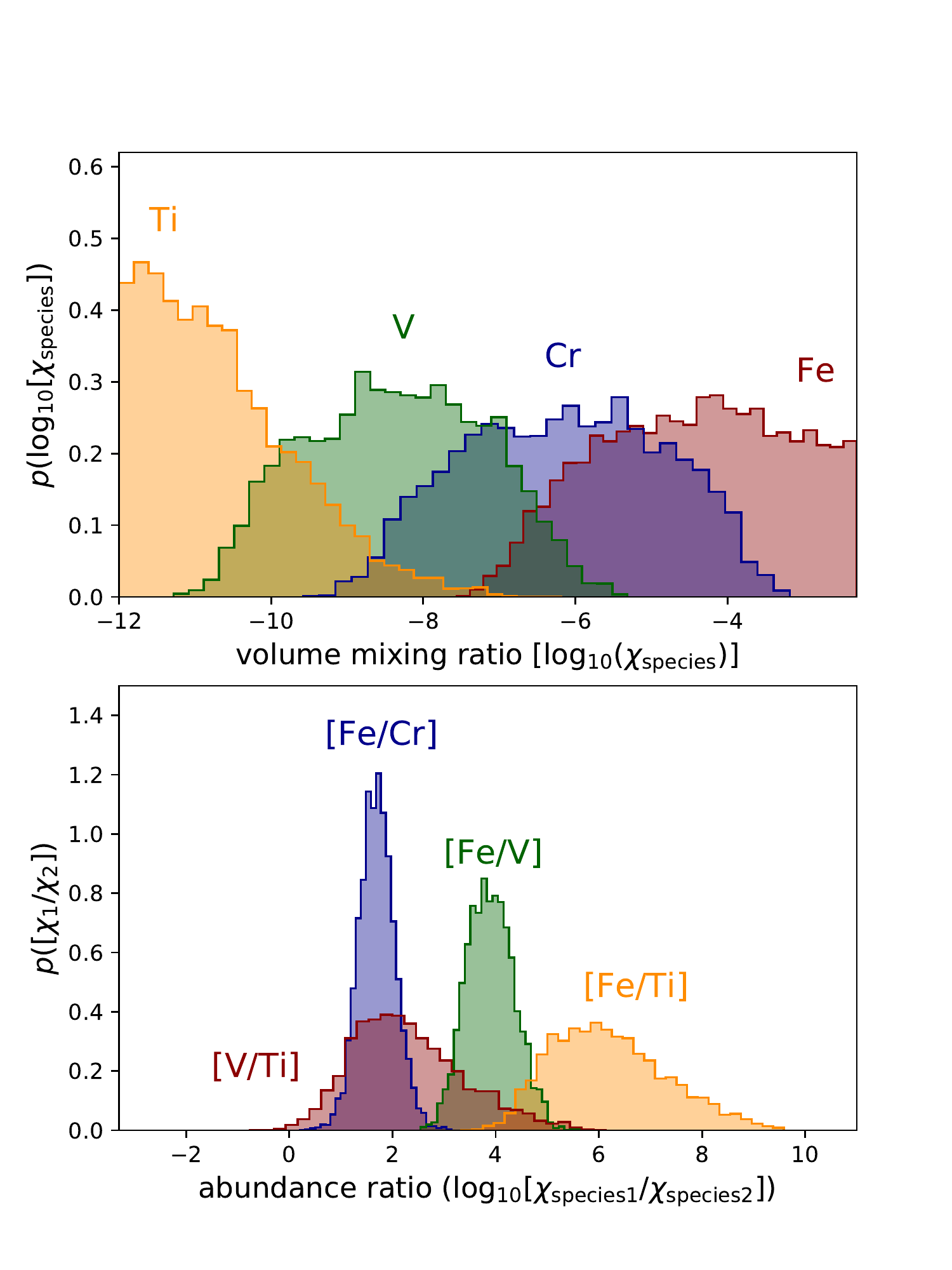}
\caption{Abundances and relative abundances from the blue retrieval identical to that shown in Fig.~\ref{fig:realdatablue_abundances} above, but with the removal of Mg from the model.}
\label{fig:realdatablue_nomg_abundances}
\end{figure}

\begin{figure}
\centering
\includegraphics[width=70mm]{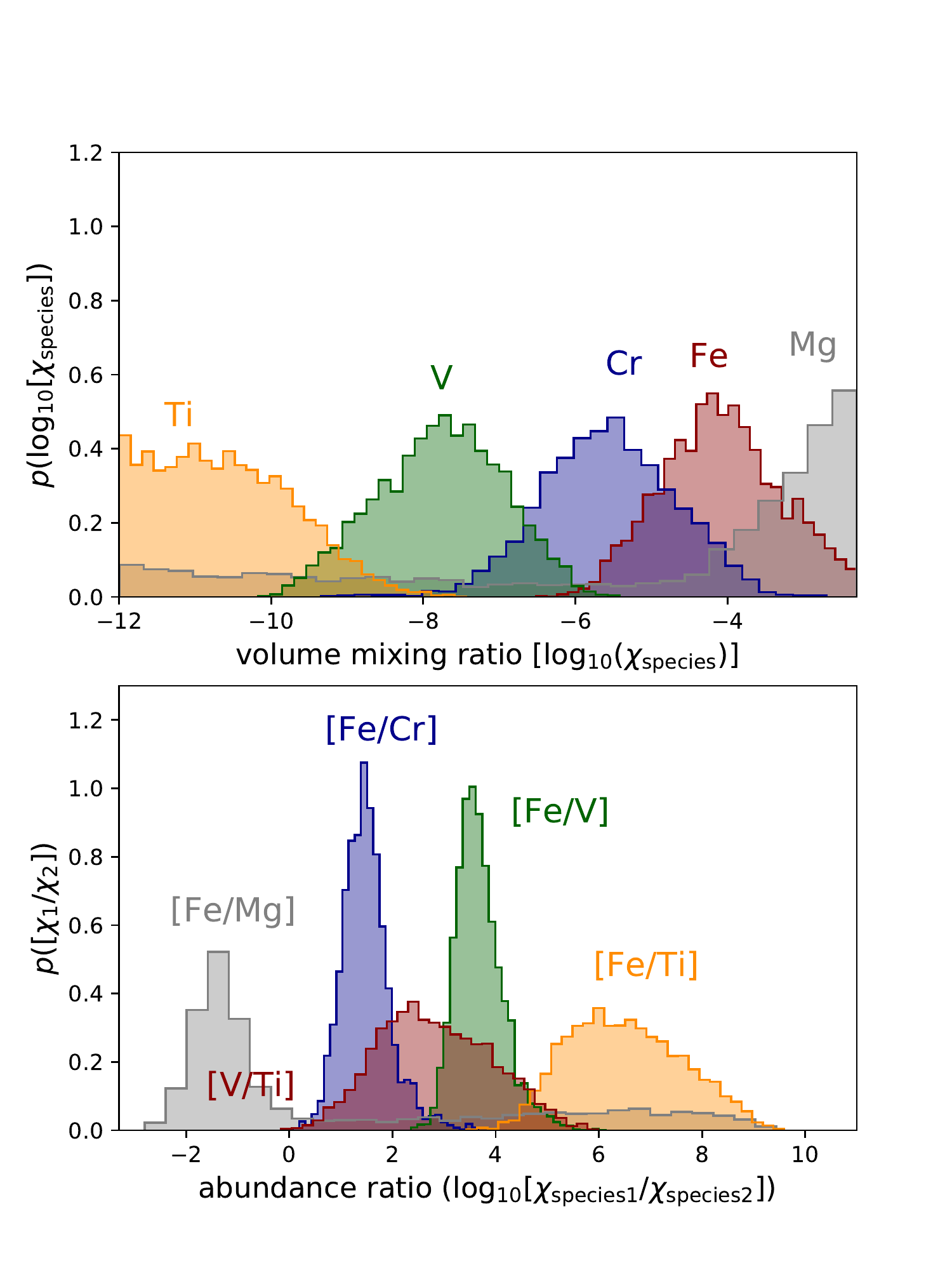}
\caption{Abundances and relative abundances for WASP-121b from the retrieval performed on the red UVES arm}
\label{fig:readdatared_abundances}
\end{figure}


\bsp	
\label{lastpage}
\end{document}